\documentclass[]{spie}  

\usepackage{lineno}
 
\usepackage{amsmath,amsfonts,amssymb}
\usepackage{bm}
\usepackage{color}
\usepackage{float}
\usepackage[colorlinks=true, allcolors=blue]{hyperref}
\usepackage[all]{hypcap}
\usepackage{placeins}
\usepackage{xcolor}
\usepackage[utf8]{inputenc}
\usepackage[T1]{fontenc}
\usepackage{mathtools}
\usepackage{enumitem}
\usepackage{graphicx}
\usepackage{listings}
\usepackage{lipsum}  
\usepackage{orcidlink}
\usepackage{tabularx}
\usepackage[encapsulated]{CJK} 
\usepackage{ucs}
\usepackage{caption}
\usepackage{subcaption}

\usepackage{devanagari} 
\usepackage{wrapfig}

\providecommand{\sorthelp}[1]{} 

\newcommand\wmap{\text{WMAP}}
\newcommand\planck{\text{Planck}}

\definecolor{citecolor}{rgb}{0.08,0.30,0.85}

\newcommand{\jhu}{William H. Miller III Department of Physics and Astronomy, Johns Hopkins University, 3701 San Martin Drive, Baltimore, MD 21218, USA}
\newcommand{\villanova}{Department of Physics, Villanova University, 800 Lancaster Avenue, Villanova, PA 19085, USA}
\newcommand{\goddard}{NASA Goddard Space Flight Center, 8800 Greenbelt Road, Greenbelt, MD 20771, USA}

\newcommand{\cfa}{Center for Astrophysics, Harvard \& Smithsonian, 60 Garden Street, Cambridge, MA 02138, USA}

\title{Design and characterization of a 60-cm reflective half-wave plate for the CLASS 90 GHz band telescope}

\author[1]{Rui Shi \orcidlink{0000-0001-7458-6946}}
\author[1]{Michael K.~Brewer}
\author[1]{Carol Yan Yan Chan \orcidlink{0000-0001-8144-556X}}
\author[2]{David T.~Chuss \orcidlink{0000-0003-0016-0533}}
\author[1]{Jullianna Denes~Couto \orcidlink{0000-0002-0552-3754}}
\author[1]{Joseph R.~Eimer \orcidlink{0000-0001-6976-180X}}
\author[1]{John Karakla}
\author[1]{Koji Shukawa \orcidlink{0000-0002-2798-2943}}
\author[1]{Deniz A. N. Valle \orcidlink{0000-0003-3487-2811}}

\author[1]{John W.~Appel \orcidlink{0000-0002-8412-630X}}
\author[1]{Charles L.~Bennett \orcidlink{0000-0001-8839-7206}}
\author[3]{Sumit Dahal \orcidlink{0000-0002-1708-5464}}
\author[3]{Thomas~Essinger-Hileman \orcidlink{0000-0002-4782-3851}}
\author[1]{Tobias A.~Marriage \orcidlink{0000-0003-4496-6520}}
\author[4]{Matthew A.~Petroff \orcidlink{0000-0002-4436-4215}}
\author[3]{Karwan Rostem \orcidlink{0000-0003-4189-0700}}
\author[3]{Edward J.~Wollack \orcidlink{0000-0002-7567-4451}}

\affil[1]{\jhu}
\affil[2]{\villanova}
\affil[3]{\goddard}
\affil[4]{\cfa}

\authorinfo{Further author information: (Send correspondence to R. Shi)\\R. Shi: E-mail: rshi9@jhu.edu}

\begin{document} 
\maketitle

\begin{abstract}
Front-end polarization modulation enables improved polarization measurement stability by modulating the targeted signal above the low-frequency $1/f$ drifts associated with atmospheric and instrumental instabilities and diminishes the impact of instrumental polarization.
In this work, we present the design and characterization of a new 60-cm diameter Reflective Half-Wave Plate (RHWP) polarization modulator for the 90 GHz band telescope of the Cosmology Large Angular Scale Surveyor (CLASS) project.
The RHWP consists of an array of parallel wires (diameter 50~µm, 175~µm pitch) positioned 0.88~mm from an aluminum mirror. 
In lab tests, it was confirmed that the wire resonance frequency ($f_\mathrm{res}$) profile is consistent with the target, $139$~Hz$<f_\mathrm{res}<154$~Hz in the optically active region (diameter smaller than 150~mm), preventing the wire vibration during operation and reducing the RHWP deformation under the wire tension.
The mirror tilt relative to the rotating axis was controlled to be $<15''$, corresponding to an increase in beam width due to beam smearing of < $0.6''$,
negligible compared to the beam's full-width half-maximum of $36'$.
The median and 16/84th percentile of the wire--mirror separation residual was $0.048^{+0.013}_{-0.014}$~mm in the optically active region, achieving a modulation efficiency $\epsilon=96.2_{+0.5}^{-0.4}\%$ with an estimated bandpass of 34~GHz.
The angular velocity of the RHWP was maintained to an accuracy of within 0.005\% at the nominal rotation frequency (2.5~Hz).
The RHWP has been successfully integrated into the CLASS 90 GHz telescope and started taking data in June 2024, replacing the previous modulator that has been in operation since June 2018.
\end{abstract}


\section{Introduction}\label{sec:intro}  
The polarized cosmic microwave background (CMB) contains valuable information about the composition, evolution, and geometry of our universe\cite{hu1997cmb,mukhanov2005physical,dodelson2020modern} and may hold direct evidence for the theory of inflation \cite{zaldarriaga1997all,kamionkowski1997statistics,baumann2009probing,abazajian2015inflation}.
Observations from two all-sky space experiments, the Wilkinson Microwave Anisotropy Probe (\wmap) \cite{bennett2012,hinshaw13} and the \planck{} satellite \cite{planck18IV,planck18VI}, have enlightened us with enormous details of the polarized microwave sky in the past two decades.
Ground-based experiments, compared to space-borne missions, benefit from more frequent maintenance and upgrades, availability of higher angular resolution due to larger mirrors, and lower costs. 
These factors have enabled ground-based experiments to make substantial contributions to the field, particularly at intermediate and small angular scales\cite{kusaka2018results,BK-XVII23,eimer23, adachi2022improved,quijoteIV,ansarinejad2024mass}. 
However, ground-based observations are challenged by low-frequency $1/f$ variations associated with atmospheric, instrumental, and calibration drifts. These effects potentially complicate the recovery of large angular scale information from the sky.

Polarization modulation is a proven technique to improve long-time instrumental stability.
This technique usually involves manipulating the polarization of the sky signal while conserving the intensity and minimizing the mixing of the polarized and unpolarized components\cite{pisano2014development}.
For millimeter and submillimeter experiments, the popular choices for the polarization modulators include the half-wave plate (HWP) and the variable-delay polarization modulator (VPM)\cite{chus12vpm}.
The HWP has been adopted in the design of various experiments such as the Atacama B-mode Search experiment (ABS)\cite{essinger-hileman2011probing}, the Balloon-borne Large Aperture Submillimeter Telescope for Polarimetry (BLASTPol)\cite{fissel2010balloon}, the E and B Experiment (EBEX)\cite{ebex2018}, the Large-Scale Polarization Explorer (LSPE) \cite{de2012swipe}, the LiteBIRD satellite\cite{hazumi2019litebird}, MAXIPOL\cite{johnson2007maxipol}, the {\scshape Polarbear} experiment (the predecessor of the Simons Array)\cite{hill2016design}, the Polarimeter fuer bolometer Kameras (PolKa) on Atacama Pathfinder EXperiment (APEX)\cite{wiesemeyer2014submillimeter}, the Simons Observatory\cite{yamada2024simons}, 
{\scshape Spider}\cite{filippini2010spider}, etc.
The VPM has been adopted in experiments such as Hertz\cite{krejny2008hertz}, the Primordial Inflation Polarization Explorer (PIPER)\cite{lazear2014primordial}, and the Cosmology Large Angular Scale Surveyor (CLASS)\cite{essinger-hileman14spie}.

\begin{figure}
    \centering
    \includegraphics[width=0.6\linewidth]{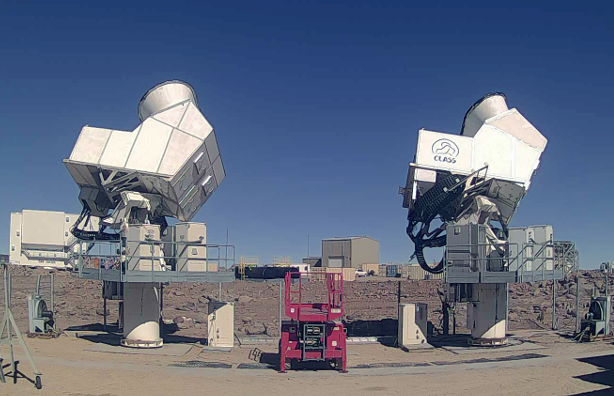}\vspace{0.5em}
    \caption{A photograph of the CLASS telescopes at Cerro Toco captured by a site camera. 
    On the left side of the image is the 150/220~GHz dual-band telescope, with an additional 90~GHz telescope to be deployed. On the right side are the 40 and 90~GHz telescopes.
    }
    \label{fig:class}
\end{figure}

The CLASS telescope array (Figure \ref{fig:class}) consists of single-frequency-band telescopes centered at 40~GHz and 90~GHz, as well as a dual-band 150/220~GHz telescope \cite{essinger-hileman14spie, harrington16spie}.
Located on Cerro Toco in the Atacama Desert of northern Chile (longitude $67^\circ$W, latitude $23^\circ$S), the 40 (90, 150/220) GHz band telescope has been operating since 2016 (2018, 2019).
A second 90 GHz band telescope is to be deployed soon.
The CLASS telescopes all adopt a similar diffraction-limited catadioptric optical design \cite{eimer12spie}, with a front-end polarization modulator as the first optical element.
Thus far, the front-end modulator has been a VPM for all the telescopes.
The VPM has significantly improved the stability of the observation \cite{harrington21,cleary22}, enabling stable measurements even for the largest angular scales ($\ell<20$)\cite{Li23,eimer23,shi24}.
As the only ground-based experiment operating with VPMs, the CLASS telescopes provide a unique platform for comparing the performance of different types of polarization modulators, a possibility that has been investigated and tested\cite{eimer2022spie}.
Building on the initial concept introduced in Eimer et al.~(2022)\cite{eimer2022spie} (hereafter E22), in this paper, we present the design and in-lab characterization of a new front-end polarization modulator, a reflective half-wave plate (RHWP), for the CLASS 90 GHz band telescope that is in operation.
The RHWP has been integrated into the CLASS 90 GHz telescope and started taking data in June 2024.

The structure of the paper is as follows: Section \ref{sec:rhwp} introduces the operational formalism and the design of the RHWP system. 
Section \ref{sec:gantry} presents the one-dimensional profilometer we designed to characterize the RHWP properties.
In Section \ref{sec:wirefres} we characterize the resonance frequency of the wire array. 
Section \ref{sec:alignment} details the alignment of the wire array and the mirror. 
Results from mechanical tests are presented in Section \ref{sec:mechanical}. 
We summarize in Section \ref{sec:conclusion}.

\section{Rotating Reflective Half-Wave Plate}\label{sec:rhwp}
A half-wave plate (HWP) modulator is designed to introduce a half-wavelength ($\pi$ radians in phase) difference between two orthogonal linear polarized states in a coordinate basis fixed to the wave plate at a particular frequency. 
The axis along which the linear polarization has the largest phase velocity is called the \emph{fast axis}, and the orthogonal counterpart is called the \emph{slow axis}.
The effect of an HWP on an incoming linearly polarized signal with polarization angle $+\alpha$ with respect to the fast axis is output by the HWP at angle $-\alpha$. 
In effect, the polarization angle has been rotated by $2\alpha$. 
A more complete formalism is described in Section \ref{ssec:formalism}. 

Taking advantage of this effect, the HWP has become a standard tool for polarization modulation. 
Among many possibilities\cite{pisano2014development}, common examples include single crystal birefringent plates\cite{abs14}, multi-layer broad-band wave plate assemblies\cite{simons19whitepaper, litebird22}, and meta-material embedded dielectric sheets\cite{pisano2016multi}. 
Each method of producing a wave plate offers unique features, making them attractive for polarization modulation.

In this work, we have designed an HWP operating in reflection constructed of an array of wires in front of a fixed-distance mirror. 
This RHWP is notable for its straightforward construction, being suitable for 60 cm (or larger) front-end modulation applications, utilizing an open geometry realized with low emissivity materials, and no required cooling.
The reflective design is compatible with commercial rotation drives and encoder readout systems.
In this section we describe the operational formalism of the wave plate (Section \ref{ssec:formalism}) and the mechanical design (Section \ref{ssec:mechanical}) of the RHWP system. 



\subsection{Operational Formalism}\label{ssec:formalism}

\begin{figure}
    \centering
    \includegraphics[width=\linewidth]{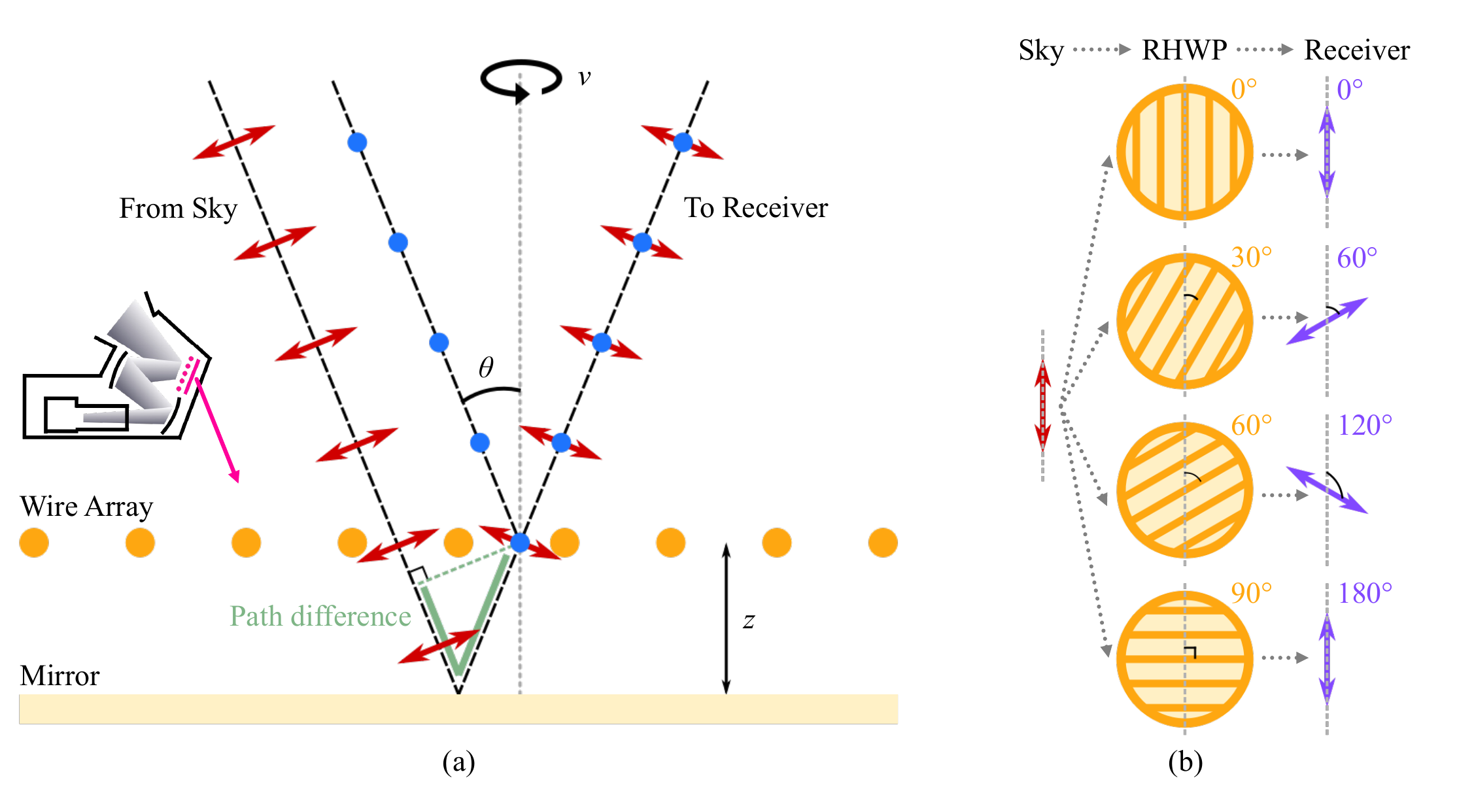}
    \caption{The model of the RHWP. 
    \textit{Panel (a)}: cross section illustrating the RHWP's operation. 
    The path difference between the two orthogonal linear polarization states induces a phase delay, enabling polarization modulation via rotation. 
    \textit{Panel (b)}: top view of the RHWP. The signal is modulated at 4$\times$ the RHWP rotation frequency.}
    \label{fig:model}
\end{figure}

The Mueller matrix of a stationary (laboratory frame) ideal phase retarder with its normal along the $z$-axis and fast axis along the $x$-axis can be expressed as:
\begin{equation}
\mathsf M_\mathrm{retarder}(\phi) = \begin{bmatrix}
1 & 0 & 0 & 0 \\
0 & 1 & 0 & 0 \\
0 & 0 & \cos\phi & -\sin\phi \\
0 & 0 & \sin\phi & \cos\phi
\end{bmatrix},\label{eq:mm_pr}
\end{equation}
where $\phi$ is the phase delay between the fast and slow axis and $\phi=\pi$ for a half-wave plate (HWP).
For an HWP with the fast axis rotated about the $z$-axis of the laboratory frame, its Mueller matrix in the laboratory frame is written as:
\begin{equation}
\mathsf M_\mathrm{HWP} = \mathsf R(-\gamma) \cdot \mathsf M_\mathrm{retarder}(\phi=\pi) \cdot \mathsf R(\gamma),\label{eq:mm_rwhp}
\end{equation}
where
\begin{equation}
\mathsf R(\gamma) = \begin{bmatrix}
1 & 0 & 0 & 0 \\
0 & \cos 2\gamma & \sin 2\gamma & 0 \\
0 & -\sin 2\gamma & \cos 2\gamma & 0 \\
0 & 0 & 0 & 1
\end{bmatrix}\label{eq:mm_rot}
\end{equation}
is the rotation matrix and $\gamma$ is the angle of rotation.
Therefore, for an ideal half-wave plate rotating with angular frequency $\omega$ (i.e., $\gamma=\omega t$), the Stokes parameters ($\vec S=[I, Q, U, V]^T$) of the outgoing light are related to the incoming ones as:
\begin{equation}
\vec S^\mathrm{out}=\mathsf M_\mathrm{HWP}\vec S^\mathrm{in}
=\begin{bmatrix}
    1 & 0 & 0 & 0 \\
    0 & \cos 4\omega t & \sin 4\omega t & 0 \\
    0 & -\sin 4\omega t & \cos 4\omega t & 0 \\
    0 & 0 & 0 & -1
    \end{bmatrix}
    \vec S^\mathrm{in},
\end{equation}
which implies that ideally, the Stokes parameter $I^\mathrm{in}$ is unchanged, $V^\mathrm{in}$ flips its sign (without modulation), and $Q^\mathrm{in}$ and $U^\mathrm{in}$ (linear polarization) are modulated at $4\times$ the rotation frequency of the HWP (Figure \ref{fig:model}, panel (b)) as:
\begin{equation}
\begin{aligned}
    Q^\mathrm{out}(t)&=\cos 4\omega t~Q^\mathrm{in} + \sin 4\omega t~U^\mathrm{in},\\
    U^\mathrm{out}(t)&=-\sin 4\omega t~Q^\mathrm{in} + \cos 4\omega t~U^\mathrm{in},
\end{aligned}
\end{equation}
and the data ($d(t)$) collected by a horizontal/vertical (perpendicular/parallel to the $x$-axis) detector are:
\begin{equation}
\begin{aligned}
    d_\mathrm H(t)&=\begin{bmatrix}
        1 & 0 & 0 & 0
    \end{bmatrix}P_\mathrm H\vec S^\mathrm{out}\\
    &=\frac 12 [I^\mathrm{in} + Q^\mathrm{out}(t)]\\
    &=\frac 12 (I^\mathrm{in} + \cos 4\omega t~Q^\mathrm{in} + \sin 4\omega t~U^\mathrm{in}),\\
    d_\mathrm V(t)&=\begin{bmatrix}
        1 & 0 & 0 & 0
    \end{bmatrix}P_\mathrm V\vec S^\mathrm{out}\\
    &=\frac12 (I^\mathrm{in} - \cos 4\omega t~Q^\mathrm{in} - \sin 4\omega t~U^\mathrm{in}),
\end{aligned}
\end{equation}
where
\begin{equation}
    P_\mathrm H=\frac12 
    \begin{bmatrix}
        1 & 1 & 0 & 0\\
        1 & 1 & 0 & 0\\
        0 & 0 & 0 & 0\\
        0 & 0 & 0 & 0
    \end{bmatrix},\qquad 
    P_\mathrm V=\frac12 
    \begin{bmatrix}
        1 & -1 & 0 & 0\\
        -1 & 1 & 0 & 0\\
        0 & 0 & 0 & 0\\
        0 & 0 & 0 & 0
    \end{bmatrix}
\end{equation}
are the Mueller matrices of horizontal and vertical polarizers.

For the CLASS project, one of the main purposes of using the HWP is to compare to the previous polarization modulator, the VPM \cite{chus12vpm,harrington18spie}, to better understand the systematics \cite{eimer2022spie}. 
Therefore, we prefer to retain the telescope's optical design, only replacing the VPM with the new modulator---meaning it needs to be reflective.
Due to the success of the VPM in improving long-term stability\cite{harrington21, cleary22}, we continue to use a wire-array-and-mirror design for the HWP. 
Figure \ref{fig:model} shows the model of the CLASS reflective half-wave plate (RHWP), which consists of a copper-plated tungsten wire array and an aluminum mirror. 
When linear polarization parallel to the wire direction reaches the RHWP, it is reflected at the wire array plane. In contrast, linear polarization perpendicular to the wire direction passes through the wire array and is reflected at the mirror surface.
This path difference between the two orthogonal linear polarization states results in a phase delay.
For light with wavelength $\lambda\gg 2r$ where $r$ is the wire radius, the geometric approximation of the phase delay can be expressed as \cite{chus12vpm}:
\begin{equation}
    \phi\approx\frac{4\pi z}{\lambda}\cos\theta,
\end{equation}
where $\theta$ is the incidence angle, and $z$ is the distance between the wire array and the mirror (or, the wire--mirror separation as in Section \ref{ssec:WMS}).
For the CLASS 90~GHz band telescope, the incidence angle is $\theta=22.2^\circ$ \cite{harrington18spie, eimer2022spie}, the band central frequency is 92~GHz\cite{dahal22}, and the phase delay for the band center is $\pi$ when $z=0.88$~mm.
The nominal rotation frequency of the RHWP was chosen to be $2.5$~Hz \cite{eimer2022spie}, so the signal is modulated at $10$~Hz.

\subsection{Design}\label{ssec:mechanical}
\begin{figure}
    \centering
    \includegraphics[width=\linewidth]{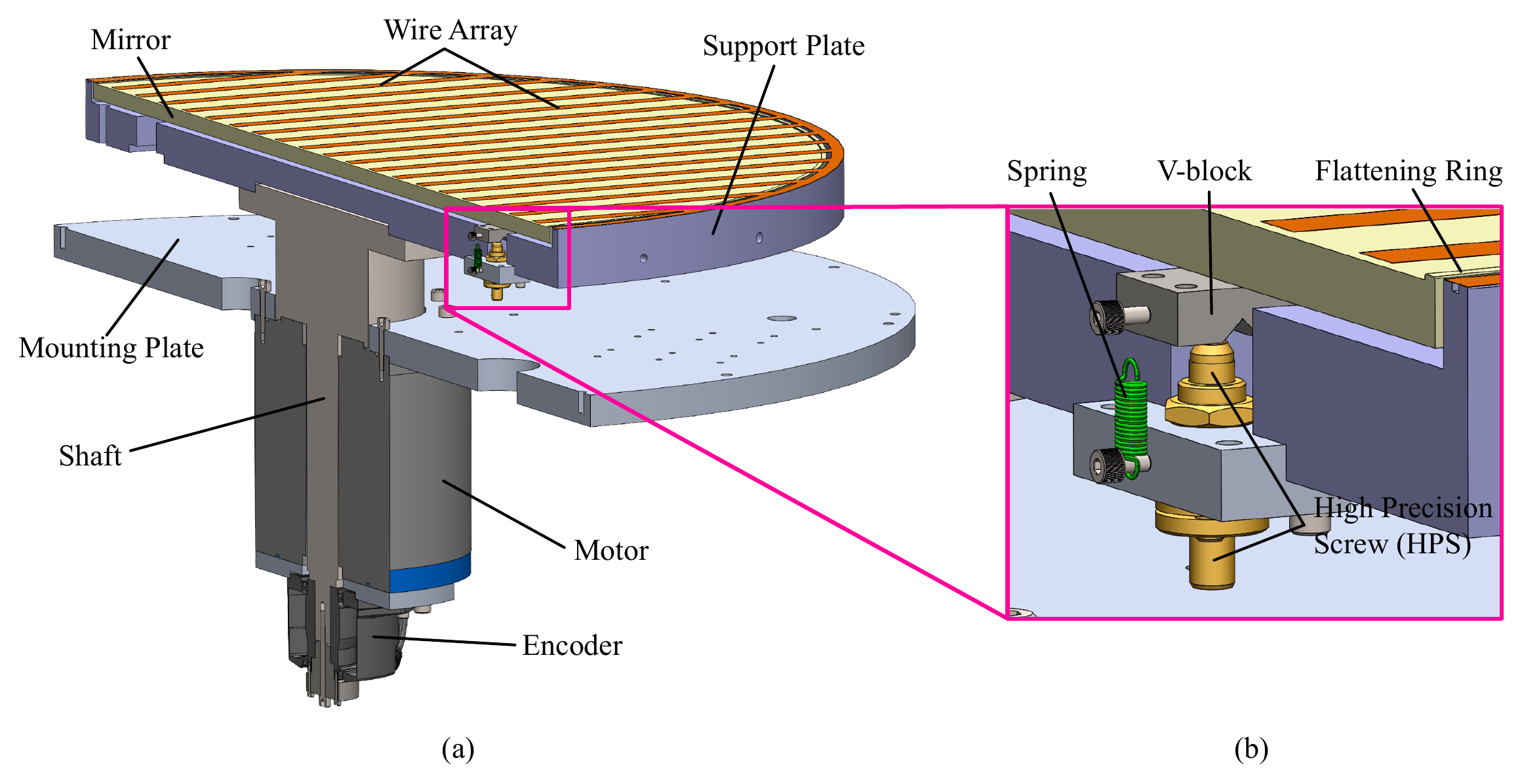}
    \caption{The cross-section view of the RHWP system. 
    \textit{Panel (a)}: The wire array, attached to the outer rim of the support plate, is schematically illustrated with a sheet of thin strips over the mirror which is captured between the back of the support plate and the wire array.
    The actual wires are cylindrical and would not be visible at this scale. 
    The RHWP and the encoder are coupled to the motor through a shaft. 
    The motor is bolted on the mounting plate which connects to the telescope frame.
    \textit{Panel (b)}: A detailed view of the holding mechanism for the mirror. 
    Three High Precision Screws (HPSs), mounted on the support plate, push against the v-grooves of the v-blocks, supporting the mirror's weight. 
    They also adjust the mirror tilt and the wire--mirror separation (Section \ref{sec:alignment}). 
    The extension springs hold the v-blocks and the HPS mounting blocks together.
    The flattening ring is part of the mirror and sets the distance between the wire and the mirror.}
    \label{fig:RHWP}
\end{figure}

Figure \ref{fig:RHWP} provides a cross-sectional view of the final version of the RHWP system.
The wire array and the mirror form the polarization modulator, both held by a support plate. 
A single shaft couples the support plate and the encoder to the motor, which is mounted on an aluminum mounting plate connected to the telescope frame.
In the following subsections, we briefly introduce the design of the main components of the RHWP system and present any major design upgrades since E22 \cite{eimer2022spie}.
\subsubsection{Wire array}\label{sssec:wire}
The specifications and manufacturing technique for the wire array are based on the fabrication of the VPM grids for CLASS  \cite{harrington18spie}. 
The manufacturing technique is adapted from earlier efforts \cite{novak1989a, Voellmer2008, Chuss2014}. 
The wire array consists of $\sim$50~$\mu$m tungsten wire coated with 2~$\mu$m thick copper using a titanium adhesion layer. 
The nominal spacing is $\sim$175~$\mu$m, chosen to roughly optimize the transmission for the radiation polarized perpendicular to the wires and the reflection for radiation polarization parallel to the wires \cite{chus12vpm}. 
The grid was manufactured using a similar technique to the CLASS VPMs, employing a cylindrical mandrel with recessed bars.
Grooves for the wires were cut into the bars using a modified CNC machine. 
The wire is then wrapped over the mandrel; the separation is set by the grooves. 
Once the wrapping is complete, the wires are epoxied to the grooved bars using Stycast 2850FT (with Catalyst 23 LV), cut, and unspooled from the mandrel.
Further details of the wire fabrication specifically for the HWP can be found in E22 \cite{eimer2022spie}.

After unspooling from the mandrel, the grooved bars were transferred to a stretching frame (Section \ref{ssec:stretch}) designed to tension the wires according to a desired profile.
Finally, the wires were epoxied (Stycast 2850FT with Catalyst 23 LV) to the top surface of the support plate rim and cut from the grooved bars.

\subsubsection{Mirror}\label{sssec:mirror}
The mirror was designed to have three critical features:\\[-20pt]
\begin{enumerate}
    \item A flat circular surface with a diameter of $\sim$ 59~cm to reflect the incoming light.\\[-18pt]
    \item A flattening ring to ensure a uniform distance between the wire array and the flat surface. The flattening ring was designed to have a height of 0.88~mm and a width of 0.125~inch.\\[-18pt]
    \item Dowel holes to locate the v-blocks (Figure \ref{fig:RHWP}, panel (b), more description can be found in Section \ref{sssec:supportplate}). There are three groups of four non-penetrating slip-fit dowel holes on the back of the mirror. 
\end{enumerate}
~\\[-36pt]

The mirror also needs to have a relatively low density but be stiff enough to bear the pressure from the wires when pushing the flattening ring against the wire array.
We used Aluminum 6061 T6 as the material for the mirror, with a thickness of 0.4~inch.
The mass of the mirror is $\sim$7~kg.

\subsubsection{Support plate}\label{sssec:supportplate}
The support plate is also made of aluminum and has three main functions:\\[-20pt]
\begin{enumerate}
    \item Provide a surface for the wires to be epoxied on and be stiff enough to not deform much under the tension from the wires.\\[-18pt]
    \item Hold the mirror.\\[-18pt]
    \item Couple to the shaft.
\end{enumerate}
~\\[-36pt]

The final version of the support plate design has been simplified from that introduced in E22\cite{eimer2022spie} to reduce lead time and cost. 
It includes a cylindrical wall on whose top surface the wires are epoxied, and a groove to increase the contact area of the epoxy.
The back of the support plate has six windows, three of which are used to mount the high-precision screws (HPSs), and the remaining three windows provide access to monitor the back of the mirror.
Three HPSs are used to hold and adjust the mirror position, each mounted on an aluminum block bolted to the support plate. 
They are evenly distributed on a ring with a 9-inch radius, centering on the support plate.
We chose the Newport AJS127-0.5H for its sensitivity (0.56~µm/$^\circ$), the axial load capacity (90~N), and the locking feature.
The top of the high-precision screw pushes against the v-grooves of the v-blocks made of AISI 304 stainless steel.
The v-blocks have the same dowel hole pattern as those on the back of the mirror and are epoxied (Stycast 2850FT with Catalyst 23 LV) on the back of the mirror, located with dowel pins.
The v-blocks and the HPS mounting blocks are held together by extension springs.
We used the Lee Spring LEM063B 02 M, for its appropriate size and tension, providing strong enough tension to hold the whole mirror without exceeding the load capacity of the HPSs. 
The tension provided by the six extension springs at the ideal position is $\sim$9~kg.
The combination of the HPSs, v-blocks, and springs hold the mirror tightly to the support plate and allows for alignment adjustments.
A circular indentation on the bottom of the support plate is used for mounting purposes (coupling to the shaft, Section \ref{sssec:shaft}). 
The mass of the support plate is $\sim$18~kg.

As in E22 \cite{eimer2022spie}, finite element analysis (FEA) simulations for the final RHWP design were conducted to understand the deformation of the support plate and eventually the mirror at different temperatures under the tension from the wire array (Section \ref{sec:wirefres}).
The deformation scales for the support plate ($\lesssim$ 80~µm) and the mirror ($\lesssim$ 30~µm across the whole surface) were similar to the results in E22 \cite{eimer2022spie}, meaning that the mirror final surface is likely to be dominated by machining tolerances (expected to be $< \pm 20~$µm).

\subsubsection{Shaft}\label{sssec:shaft}
The coupling between the support plate, the encoder, and the motor has been simplified compared to the design in E22 \cite{eimer2022spie}.
A single shaft couples the support plate and the encoder to the motor.
The shaft is made of AISI 304 stainless steel to match the requirement for the coefficient of thermal expansion (CTE) of the encoder bore.
The top cylinder of the shaft was used for the concentric alignment between the shaft and the support plate.
As a $\sim40^\circ$C ($-20^\circ$C to $20^\circ$C) thermal cycle is expected to happen for these parts, a 4~mil gap in diameter was left to accommodate the CTE mismatch between the support plate and the shaft.
A confocal distance sensor (Section \ref{sec:gantry}) was used for the center alignment of the support plate and the shaft, and an off-center of $<5$~µm (radially) can be achieved.
The mass of the shaft is $\sim$11~kg.

\subsubsection{Drive system}\label{sssec:drive}
The servo system consists of a 16-pole Kollmorgen DH062M-12-1310 direct drive AC brushless servo motor, a Heidenhain RCN 2581 angle encoder, and an ABB Microflex e100 servo drive. 
The servo drive has an electromagnetic interference filter with $\sim$250~kHz cutoff on its motor power output to reduce the risk of interference from pulse width modulation motor drive currents. 
The encoder has an accuracy of $2''$ and features two encoder readout signals. 
A serial Endat signal with 28-bit resolution and one-volt peak-to-peak sine/cosine analog signals with 16384 cycles per revolution. 
The Endat readout is used for recording the position, while the analog signal is sent to the servo drive for feedback in the servo loop. 
The servo drive's position and velocity loops run at 4~kHz and upsample the encoder signal by a factor of 16 to 262144 counts per revolution, yielding the capability of positioning the RHWP to a resolution of $5''$.

The controlling computer is a PICMG 1.3 single-board computer running the VxWorks real-time operating system mounted in a 13-slot mixed PCI/PCI Express passive backplane. 
This computer runs its own servo loop at 20 Hz and features two control modes: position and velocity. 
In the normal velocity mode, the computer simply sends constant velocity jog commands to the servo drive over a TCP/IP connection along with a watchdog signal that is used by the servo drive to cease motion if the connection is lost. 
It also reads the current position and velocity from the drive for use in the control loop along with various quantities such as bus voltage, motor current, and following error, which are then recorded for diagnostic purposes. 
Testing has shown that the RHWP maintains a constant rotation frequency of 2.5~Hz to an accuracy of within 0.005\% (Section \ref{sec:mechanical}). 

The controlling computer also records the position of the RHWP. 
This is done in an interrupt handler, which receives interrupts along with associated 32-bit consecutive serial numbers from a sync box running at $\sim$201~Hz. 
The interrupt handler reads the Endat serial position from the encoder using an Addi-Data APCIe-1711 PCI Express card. 
It also reads a time stamp from a Symmetricom bc637PCI-V2 GPS clock card, which has a resolution of 100~nanoseconds. 
Both are recorded to disk along with the serial number for each interrupt. 
The same sync box sends its interrupts and serial numbers to the multi-channel electronics, which record the detector data. 
The serial numbers are then used to synchronize the detector data with the position and time.

\subsubsection{Mounting structure}
\begin{figure}[t]
    \centering
    \includegraphics[width=\linewidth]{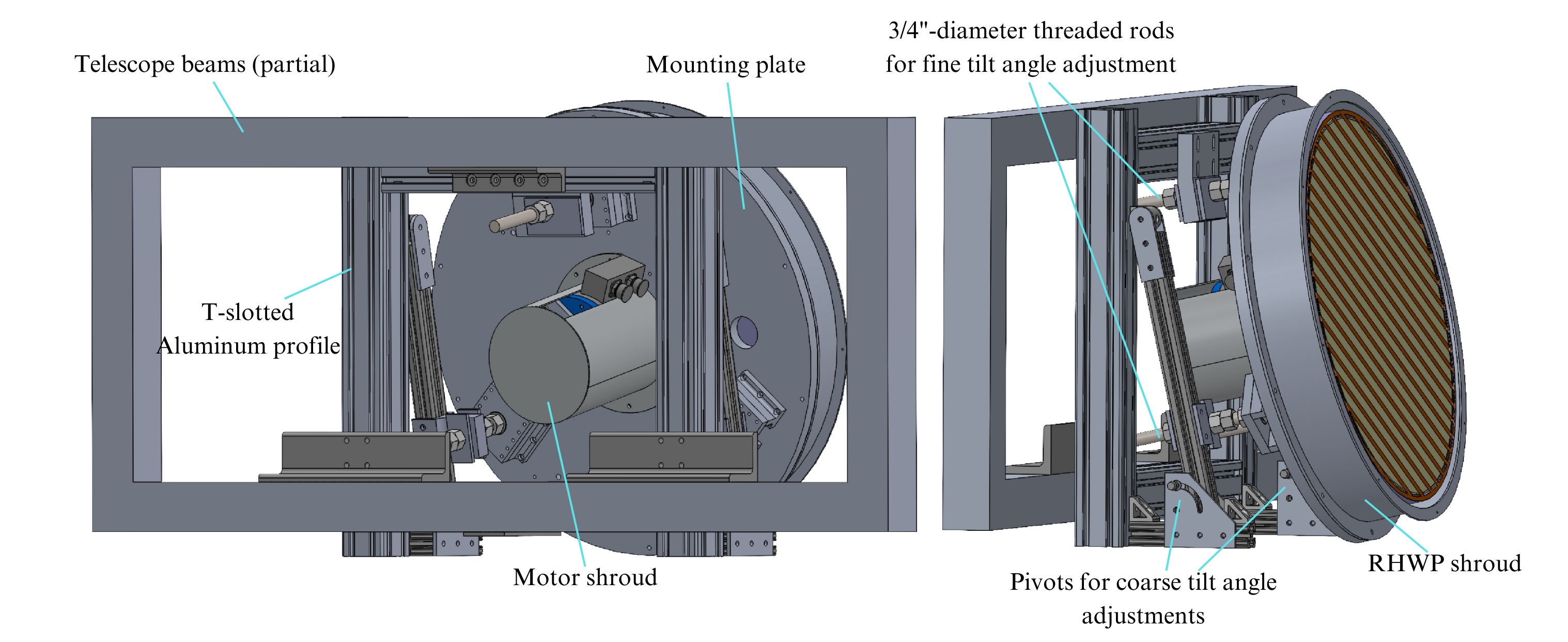}
    \caption{The mounting structure of the RHWP system. Only the part of the telescope beams to which the RHWP is attached is shown. The tilt angle of the RHWP can be coarsely adjusted by two pivots and finely adjusted by three threaded rods. }
    \label{fig:mounting}
\end{figure}

Figure \ref{fig:mounting} shows the mounting structure of the RHWP system.
The mounting plate of the RHWP system is connected to the telescope optics cage through three 3/4-inch diameter threaded rods. 
The tilt and position of the RHWP assembly can be finely tuned by adjusting the length of the threaded rods. 
Each threaded rod is attached on a spherical bearing, allowing a maximum angle of 9$^{\circ}$ between the rod and the normal of the attachment surface. 
The other end of the rod is attached to a mounting structure made of an aluminum T-slotted profile, which interfaces with the telescope beams. 
The pivot in the mounting structure allows additional elevation angle adjustment. 
The use of a T-slotted frame provides flexibility to adjust the position of the RHWP horizontally and vertically along the direction of the telescope beams. 
The motor and the RHWP are enclosed by the motor shroud and the RHWP shroud respectively, to separate the moving parts from the external environments.

\section{Profilometer}\label{sec:gantry}
\begin{figure}
\begin{minipage}{\linewidth}
    \centering
    \raisebox{-0.5\height}{\includegraphics[width=0.73\linewidth]{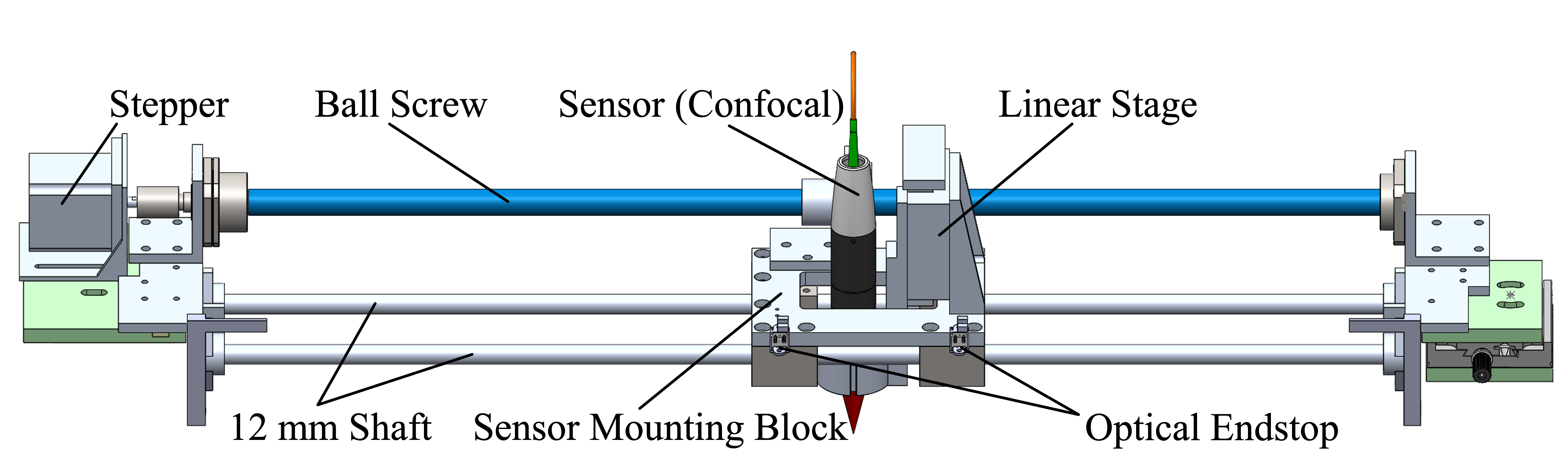}}
    \hspace*{0in}
    \raisebox{-0.52\height}{\includegraphics[width=0.25\linewidth]{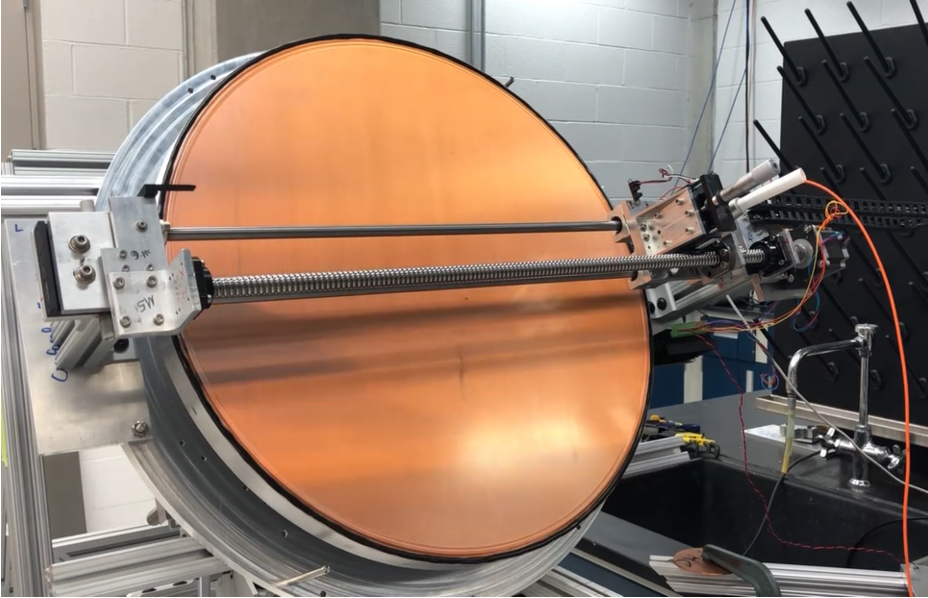}}
    \caption{\textit{Left}: The profilometer model. 
    The sensor mounting block, held by a ball screw and two 12~mm shafts, can be accurately positioned by a stepper. Either a microscope or a confocal distance sensor can be mounted on this block. 
    The linear stage enables a 25~mm sensor traveling along the vertical direction.
    The two optical endstops limit the trajectory of the sensor mounting block.
    \textit{Right}: A photograph of the profilometer mounted on the RHWP system.}\label{fig:profilometer}
\end{minipage}
\end{figure}

We employed a one-dimensional profilometer to characterize the RHWP properties (Figure \ref{fig:profilometer}).
We adapted the XY-gantry design from Harrington et al.~(2018)\cite{harrington18spie} to suit our specific requirements.
The sensor mounting block on the profilometer is supported by two 12~mm shafts and one ball screw with a 5~mm pitch, each exceeding 700~mm in effective length.
A linear stage facilitates 25~mm of vertical sensor travel.
Positioning of the sensor mounting block is achieved using a NEMA 23HS22-2804S stepper motor, controlled by an Arduino UNO R4 MINIMA board and DM254 micro-step driver.
The position repeatability was achieved within 0.04~mm.
The travel range of the sensor mounting block was limited by two optical endstops (LERDGE Optical Endstop-4001).
The profilometer was bolted on the mounting plate using two T-slotted profiles to roughly set the distance between the sensor and the RHWP (Figure \ref{fig:profilometer}).

The profilometer was used to analyze the wire resonance frequency, mirror tilt relative to the rotation axis, and wire--mirror separation.
Depending on the task, we employed either a microscope with a camera or a confocal distance sensor to scan the top surface of the RHWP.
For microscopic inspections, we utilized a setup similar to Harrington et al.~(2018)\cite{harrington18spie}: a 10$\times$ objective and a 20$\times$ wide-field eyepiece provided 200$\times$ magnification, coupled with a 2-megapixel USB camera via a macro lens.
This configuration provides a field-of-view of approximately 250~µm in diameter and a depth resolution of $\sim$10~µm, sufficient for measuring the distance between the wire and the mirror (0.88~mm).
Alternatively, we employed the IFS2406-10 confocal distance sensor with an IFC2421 controller from MICRO-EPSILON.
The confocal sensor offers a 10~mm measuring range and a dynamic resolution of 0.2~µm, suitable for simultaneous measurement of the wire array and mirror.
With a 15~µm light spot diameter and a maximum $\pm 13.5^\circ$ measuring angle, the sensor is capable of resolving the profile of individual wires (diameter $\sim$50~µm).

To scan the entire top surface of the RHWP, we integrated the one-dimensional linear motion of the profilometer with the RHWP rotation.
The sensor trajectory relative to the RHWP system was determined using a portable coordinate measuring machine (CMM)

\section{Wire Resonance Frequency}\label{sec:wirefres}
The wire resonance frequency must satisfy two constraints: it should exceed the signal band ($\sim$10~Hz) to minimize vibration during RHWP operation, and it should remain low enough to prevent excessive deformation of the support plate due to wire tension.
The relation between wire resonance frequency ($f_\mathrm{res}$) and tension ($T$) is:
\begin{equation}
T = 4\rho L^2 f_\mathrm{res}^2,
\label{eq:tension}
\end{equation}
where $L$ is wire length, and $\rho$ is the mass per unit length.
Our final choice of the resonance frequency/wire tension profile is one step away from E22\cite{eimer2022spie} based on the following consideration: the temperature of the RHWP could vary from $-20~^\circ$C to $20~^\circ$C, and the CTE mismatch between tungsten ($4.5\times 10^{-6}~\mathrm{m/m~^\circ C}$ for the wires) and aluminum ($23.4\times 10^{-6}~\mathrm{m/m~^\circ C}$ for the support plate) significantly alters the wire tension during a thermal cycle.
To match the wire tension profile at $-20~^\circ$C as in E22\cite{eimer2022spie}, the tension per wire at room temperature ($20~^\circ$C) must be increased by 0.61~N, resulting in a total wire array tension of $\sim$3120~N at room temperature (ranging 0.64~N to 1.03~N from per wire), and $f_\mathrm{res}>139$~Hz for all wires at $20~^\circ$C.

In practice, we used a custom-designed stretching frame to adjust the wire array tension and a profilometer with a confocal distance sensor to measure the resonance frequency.

\subsection{Stretching Frame}\label{ssec:stretch}



\begin{figure}
\includegraphics[width=0.732\columnwidth]{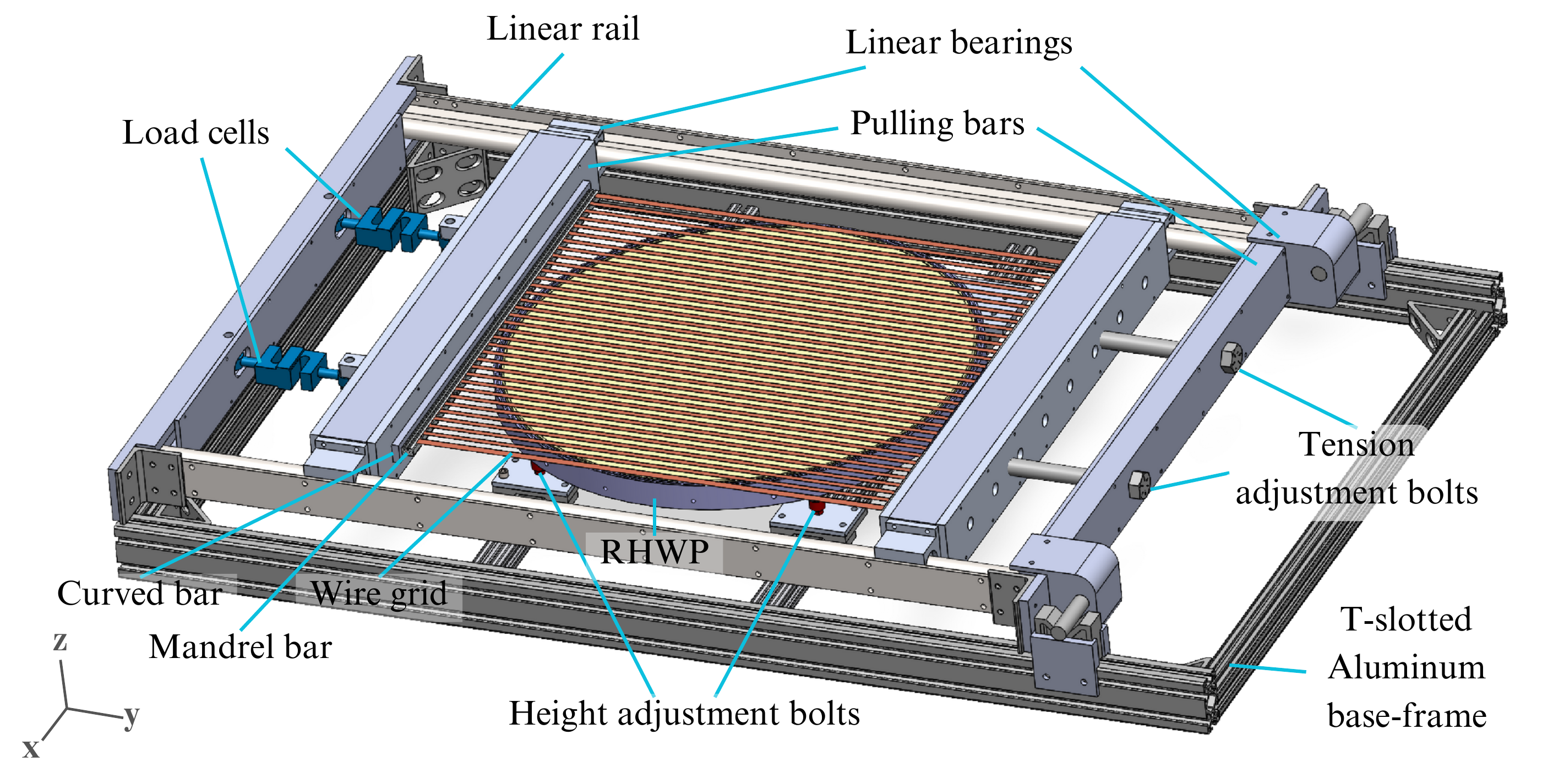}
\includegraphics[width=0.26\columnwidth]{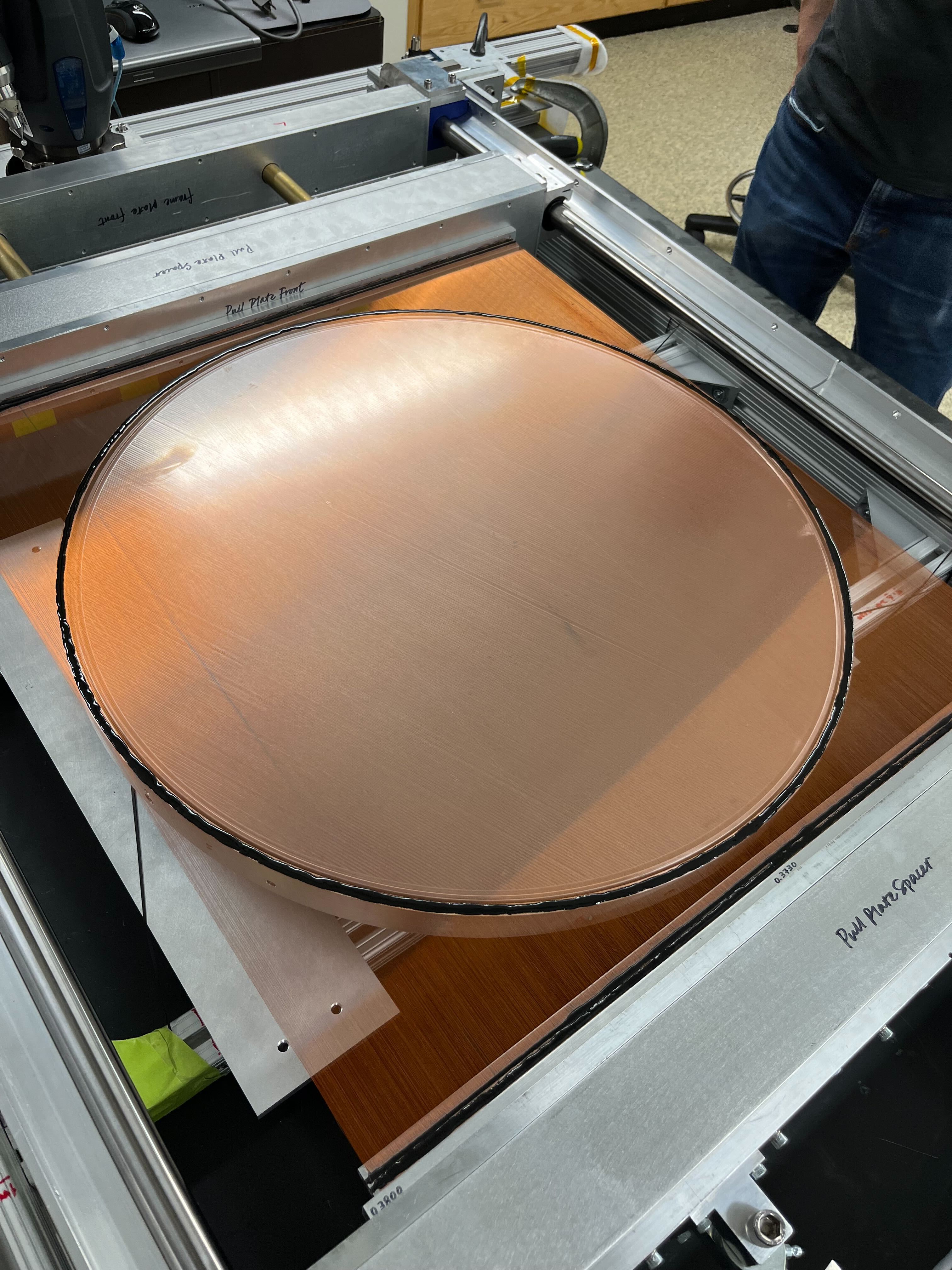}
\caption{\textit{Left}: The stretching frame model. The width of individual wires is exaggerated for visualization. \textit{Right}: the tension of the wire array is maintained by the stretching frame while the wire array is epoxied to the rim of the support plate.}
\label{fig:stretching frame}
\end{figure}

\begin{figure}
\includegraphics[width=0.45\columnwidth]{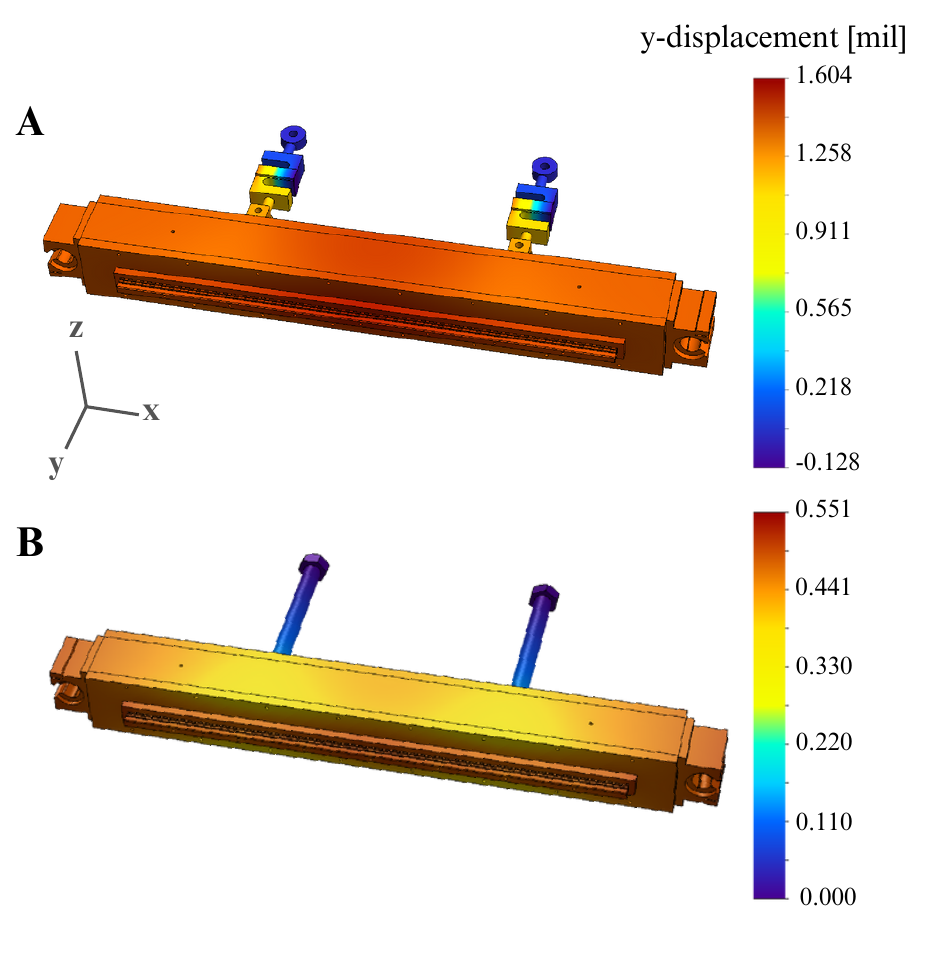}
\includegraphics[width=0.5\columnwidth]{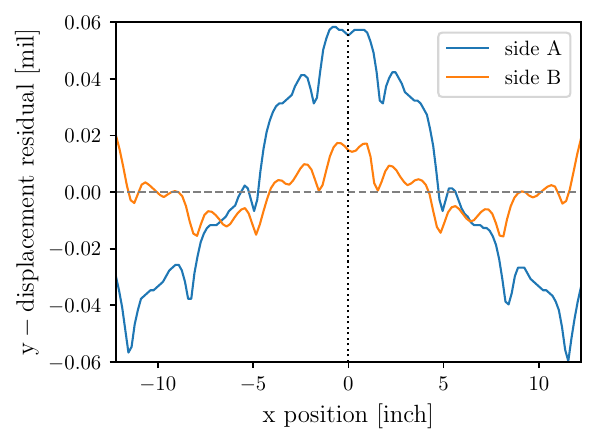}
\caption{\textit{Left}: FEA simulation results for the deformation of mandrel bars on the load cell side (A) and adjustment bolt side (B). 
\textit{Right}: the $y$-displacement of the mandrel bars on the load cell side (A, blue) and adjustment bolt side (B, orange) with the mean removed (denoted as the $y$-displacement residual). 
Direction towards the center of the RHWP is defined as positive $y$ direction for both mandrel bars.
The black dotted (gray dashed) line provides the zero $x$ position (zero $y$-displacement residual) for reference.
The deformation along the wire direction is less than 0.16~mil for both sides of the mandrel bars, which is only 2\% of the targeted 6.8~mil change between the middle and edge of the curved surface and is hence negligible.}
\label{fig:deformation}
\end{figure}

A custom-designed stretching frame was used to tension the wires to the desired profile. 
Figure \ref{fig:stretching frame} shows the model and a picture of the wire array being tensioned by the stretching frame.

The two pulling bars support the wire tension and are coupled to the rails through linear bearings.
Two curved bars were used to achieve the desired tension profile (E22 \cite{eimer2022spie} and Figure \ref{fig:tension}).
Each curved bar has a flat surface in contact with the pulling bar and a curved surface on the other side. 
The mandrel bars holding the wire array are attached to the curved bars and conform to the shape of the curved surfaces. 
The curved surface profile was chosen such that the wire tension profile matches the target as closely as possible when the curved bars are parallel to each other and support a total wire array tension of $\sim3120$~N.
Following the coordinate system defined in Figure \ref{fig:stretching frame}, the profiles of the curved surfaces are $y_{\pm}=\mp0.0436x^2+\mathrm{const}$ where $x$ is in inches, $y$ is in mil, and the subscript $_{\pm}$ indicates the curved bar in the $y>0$ or $y<0$ region.
The $y$-difference between the middle and edge for both curved bars is $\sim$6.8 mil.
The tension profile can also be finely adjusted by loosening the bolts that hold the pulling bar assemblies on each side.
One of the pull bars is attached to two fixed load cells.
The position of the other pull bar is adjustable via two 3/4-inch-diameter bolts.

FEA simulations were used to optimize the pull bar design by minimizing the static deformation of the mandrel bars when the wire array achieves the nominal tension of $\sim3120$~N, because if the mandrel bars deform significantly, the wire array tension profile deviates from the target.
Simulation results (Figure \ref{fig:deformation}) show that for the final design, the mandrel bar deformation along the wire direction is less than 0.16~mil at both sides, which is only 2\% of the 6.8~mil $y$-difference target and therefore negligible. 

After the wires were stretched to the desired profile by the stretching frame, the support plate was brought into contact with the wire array by three height adjustment bolts beneath it. 
The wire array was then permanently secured to the rim of the support plate with epoxy (Figure \ref{fig:stretching frame}).

\subsection{Measurements}\label{ssec:wirefres}
The method for measuring the wire resonance frequency is straightforward:~\\[-20pt]
\begin{enumerate}
    \item A speaker was used to excite the wires, with a function generator (Stanford Research DS340) controlling the speaker to perform phase-continuous linear frequency sweeps across a specific frequency range (single direction from low to high).~\\[-18pt]
    \item The confocal distance sensor was focused on a single wire, and the confocal distance data were recorded with and without excitation for a certain period.~\\[-18pt]
    \item The periodogram of the distance data was analyzed by comparing the data with and without excitation to identify any significant spikes.
    The frequency of the lowest significant spike was identified as the resonance frequency of the wire.
\end{enumerate}
~\\[-36pt]
\begin{figure}
    \centering
    \includegraphics[width=\linewidth]{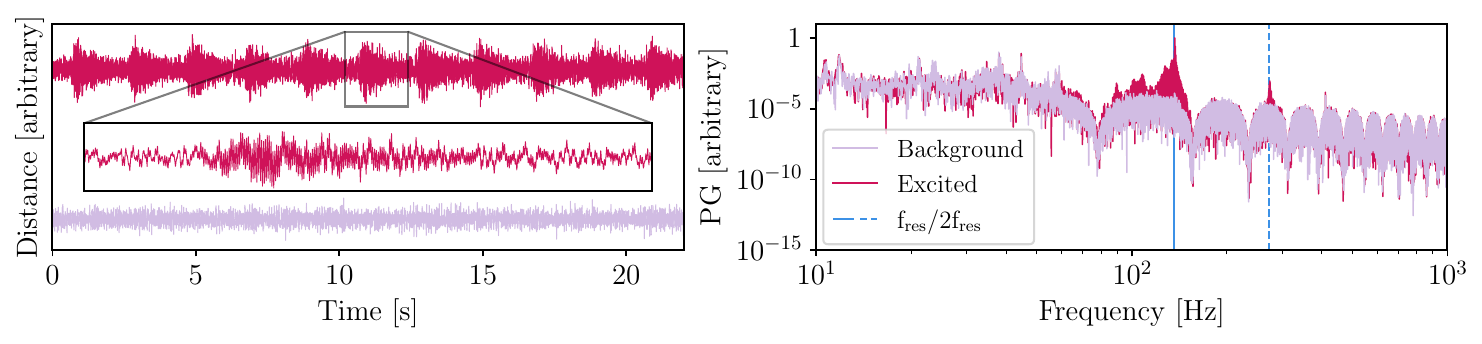}
    \caption{\textit{Left}: the confocal distance sensor measurements when the wires are excited (red) and not (purple), details about the measurements can be found in the text.
    The data was shifted along the $y$-axis, and a high-pass filter (> 5~Hz) was applied for visualization purposes. 
    \textit{Right}: the periodogram (PG) of the data on the left panel, normalized to unity at the peak of the excited case.
    The first and second harmonics (blue solid and dashed) can be identified.}
    \label{fig:fres}
\end{figure}
\begin{figure}
    \centering
    \includegraphics[width=\linewidth]{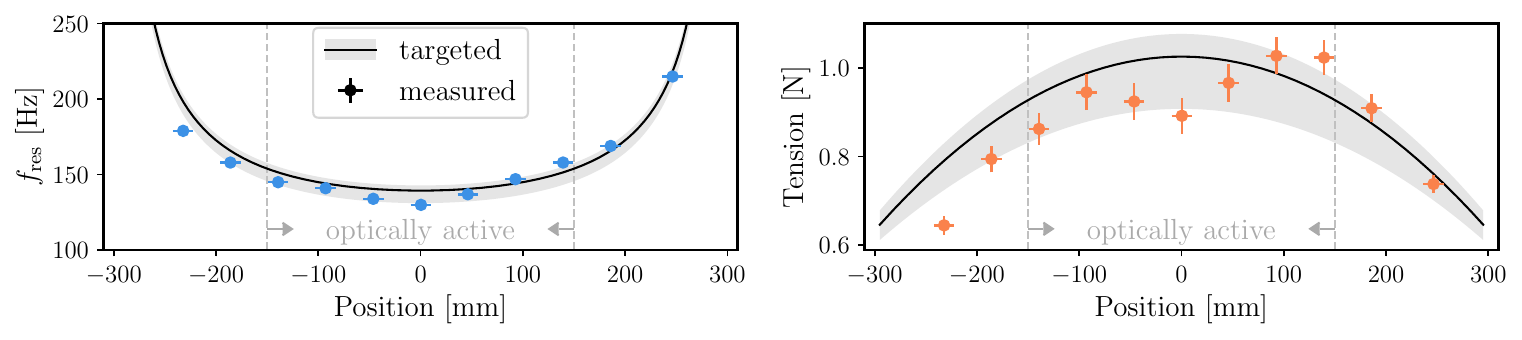}
    \caption{\textit{Left}: the measured resonance frequency profile (blue), which is mostly consistent with the target (black solid with gray region). The optically active (diameter <30~cm) region is marked with gray dashed lines.
    \textit{Right}: the similar information for the tension.}
    \label{fig:tension}
\end{figure}

During data acquisition, the frequency sweep range changed from position to position.
It was chosen such that the lowest frequency was always lower than half of the identified resonance frequency, ensuring that higher harmonics of the fundamental mode were not misidentified as the resonance frequency.
The period of the frequency sweep was $\sim$2$-$3 seconds depending on the frequency range.
A Hann window function was applied when computing the periodogram.


Due to time constraints, tension profile adjustments before epoxying the wires onto the support plate were based on a few ($\sim$6 per iteration) measurements evenly distributed across the wire array.
During these measurements, the speaker was placed on the pulling bar closer to the load cells.
More detailed measurements were conducted after the RHWP was fully assembled.
The left panel of Figure \ref{fig:fres} shows the recorded distance of one wire without excitation (background) and with excitation. 
The wire vibration amplitude increased significantly as the speaker frequency swept across the resonance frequency.
The periodogram of the excited case (Figure \ref{fig:fres}, right panel) shows spikes around 136.5~Hz and 273~Hz; the former is identified as the resonance frequency for that wire.

Figure \ref{fig:tension} compares the measured resonance frequency profile (also converted to tension) to the targeted profile, and they are generally consistent with each other.
The measured resonance frequency profile is also consistent with the two load cell readings.
The load cells were not fully relied upon for adjusting the tension profile due to relatively large calibration uncertainty ($\sim$10\%) and coarse spatial resolution.

%


\section{Wire--Mirror Alignment}\label{sec:alignment}
In this section, we present the in-lab alignment results, including the mirror tilt relative to the rotation axis, and the wire--mirror separation.
The in-lab alignment was performed to validate the alignment methods, summarize the procedure, and prepare data pipelines.
For safety, the mirror was sitting on the bottom of the support plate during transportation, and the alignment work was redone at the site.

\subsection{Mirror Tilt}\label{ssec:tilt}
Our current requirement on the mirror tilt relative to the rotation axis (or, the mirror wobbling for short) is based on its impact on the beam: the observed beam map ($B^\mathrm{obs}(\theta_x,\theta_y)$) is equivalent to the instrumental beam ($B^\mathrm{inst}(\theta_x,\theta_y)$) convolved with the trajectory of the mirror wobbling ($K(\theta_x,\theta_y)$) as:
\begin{equation}
\begin{aligned}
    B^{\mathrm{obs}}(\theta_x,\theta_y)&=\int B^\mathrm{inst}(\theta_x^\prime,\theta_y^\prime)K(\theta_x-\theta_x^\prime,\theta_y-\theta_y^\prime)~d\theta_x^{\prime}d\theta_y^{\prime},
\end{aligned}
\end{equation}
where $\theta_\mathrm{wobb}$ is the angle between the mirror's normal direction and the rotation axis.
We denote the beam smearing effect as $\Delta_\mathrm{FHWM}$, which is the difference between the full-width half maximum (FWHM) of $B^\mathrm{obs}$ and $B^\mathrm{inst}$. 
Assuming a perfectly flat mirror, the convolution kernel can be expressed as:
\begin{equation}
    K(\theta_x,\theta_y)=\delta^{(2)}[\theta_x^2+\theta_y^2-(2\theta_\mathrm{wobb})^2],
\end{equation}
where $\delta^{(2)}$ is the two-dimension Dirac delta function.

The FWHM of the CLASS 90 GHz telescope is $36'$ \cite{datta2023class}, with a nominal pointing tolerance of $2'$.
Practically, we can easily achieve $\theta_\mathrm{wobb}<15''$, which corresponds to a negligible $\Delta_\mathrm{FWHM}=0.6''$.

Other systematic effects, including the T-to-P leakage, require more careful modeling and will be considered in future publications.

\subsubsection{Measurements}

We used the confocal distance sensor to map the mirror--sensor distance and estimate $\theta_\mathrm{wobb}$ from it. 
The measurements were performed as follows:\\[-20pt]
\begin{enumerate}
    \item Position the sensor at a desired location with the stepper.\\[-18pt]
    \item Rotate the mirror at a constant speed of 60 $^\circ$/s while keeping the sensor location fixed.\\[-18pt]
    \item Record the sensor distance and RHWP angle encoder data for one minute (10 periods).\\[-18pt]
    \item Repeat steps 1--3 until data from all sensor locations are collected.
\end{enumerate}
~\\[-18pt]
For each sensor location, we will refer to the resulting mean-subtracted data as the ``sensor distance residual''.

\begin{figure}
    \centering
    \includegraphics[width=\linewidth]{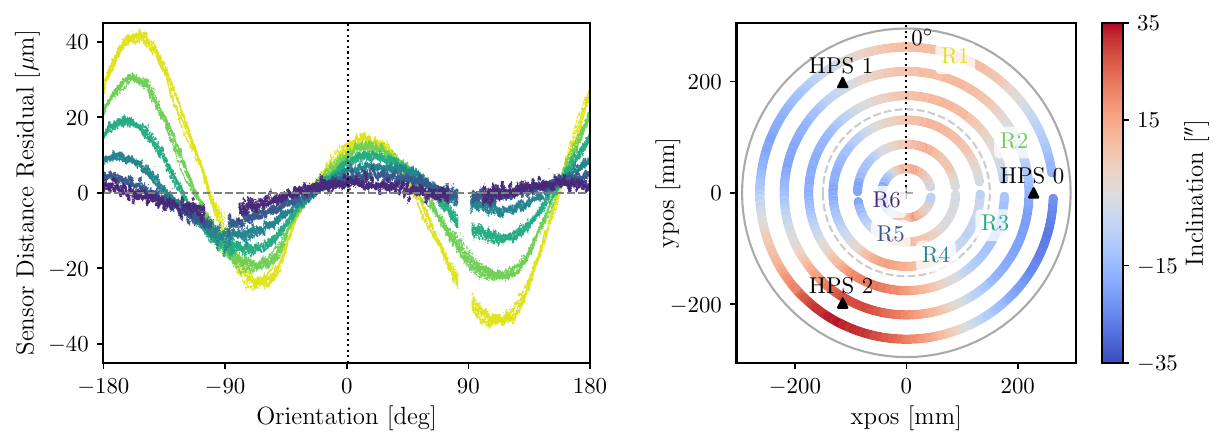}
    \caption{\textit{Left}: the sensor distance residual measurements at different sensor locations after the final HPS tilt adjustments. The black dotted and gray dashed lines mark the zero orientation angle and zero residual, respectively. The colors correspond to the R1--R6 labels on the right panel.
    \textit{Right}: the inclination at different orientations per sensor location, annotated from R1 to R6. 
    The black dotted line marks the $0^\circ$ orientation angle, which increases clockwise. The locations of the three HPSs where adjustments were made are marked with black triangles. 
    The fact that the sensor distance residual and the inclination are dominated by the component with a $180^\circ$ period instead of a $360^\circ$ one, indicates that the mirror--sensor distance change is dominated by the mirror flatness imperfections rather than tilt.}
    \label{fig:tilt}
\end{figure}

\begin{table}
\caption{Properties for the mirror wobbling measurements. $\theta_\mathrm{wobb}$ at all sensor locations was controlled to be smaller than $15''$, which corresponds to a beam smearing amplitude smaller than a negligible $0.6''$.} 
\label{tab:tilt}
\begin{center}       
\begin{tabularx}{.7\textwidth}{|r|X|X|X|X|X|X|}
\hline
\rule[-1ex]{0pt}{3.5ex}   & R1 & R2 & R3 & R4 & R5 & R6  \\
\hline
\rule[-1ex]{0pt}{3.5ex}  Radius [mm] & 262 & 218 & 175 & 131 & 88 & 44  \\
\hline
\rule[-1ex]{0pt}{3.5ex}  $\theta_\mathrm{wobb}~['']$ & $9.8$ & $5.8$ & $3.2$ & $5.8$ & $7.9$ & $9.5$ \\
\hline 
\rule[-1ex]{0pt}{3.5ex}  $\Delta\theta_\mathrm{wobb}~['']$ & $0.8$ & $0.9$ & $1.2$ & $1.6$ & $2.4$ & $4.7$ \\
\hline 
\rule[-1ex]{0pt}{3.5ex}  Tilt Orientation [$^\circ$] & $79$ & $87$ & $99$ & $151$ & $-141$ & $-127$ \\
\hline 
\end{tabularx}
\end{center}
\end{table}

The profilometer is not perfectly rigid, and we noticed that the vertical position of the sensor at different locations changes significantly and found it hard to trace. 
Therefore, we analyzed the sensor distance residual data at each location independently.
We obtained the relation between the sensor distance residual and the encoder position by interpolating the sensor distance residual to the encoder timestamps, due to the different sampling rates of the sensor (10~kHz) and the encoder (500~Hz). 
The RHWP rotated at 60~$^\circ/s$, and the encoder sampling rate of 500~Hz provides an angular resolution of $7.2'$, which is sufficient for this task.
Finally, using CMM measurements, we converted the encoder data and the sensor location into mirror $x$ and $y$ positions at each timestamp and mapped the sensor distance residual data.
The normal of the best-fit plane of the sensor distance residual data at each radius provides the tilt orientation at that radius, and the angle between it and the $z$-axis was interpreted as $\theta_\mathrm{wobb}$ for each sensor location.
During mirror tilt alignment, we iterate between taking measurements (the tilt orientation and $\theta_\mathrm{wobb}$) and adjusting the HPSs until satisfied.

The left panel of Figure \ref{fig:tilt} visualizes the relationship between the sensor distance residual and orientation for the final in-lab measurements.
Some data were excluded from this analysis because when the wires are perpendicular to the direction of the connection between the sensor and the mirror center, the presence of the wire has a non-negligible influence on the sensor reading.
We found that the sensor distance residual is dominated by the component with a $180^\circ$ period instead of a $360^\circ$ one, indicating that the mirror-sensor distance change is dominated by the mirror flatness imperfections rather than tilt.
In the right panel of Figure \ref{fig:tilt}, we present the ratio of the sensor distance residual and the rotation radius at each sensor location. 
Note that this reflects the inclination at different orientations and differs from $\theta_\mathrm{wobb}$, which characterizes the overall tilt at each radius.
The quadrupole pattern is consistent with the curves in the left panel, and the inclination scales at different radii are similar.

Table \ref{tab:tilt} lists the measurement properties. 
The uncertainty on the wobble, $\Delta\theta_\mathrm{wobb}$ was propagated from the uncertainty in the sensor reading ($\Delta d$).
We estimated $\Delta d$ at each sensor location through iterative fitting.
In each iteration, we first fit a model $\eta=\mathcal A\sin(2\pi \xi/T+\xi_0)$ where $\eta$ is the residual from the previous iteration (starting from the sensor distance residual) and $\xi$ is the orientation. 
The period $T$ is fixed to be $360^\circ/i$ where $i$ is the fitting iteration index. The amplitude $\mathcal A$ and phase $\xi_0$ are the fitting parameters.
We then subtracted the best-fit model from the current residual and calculated the standard deviation of the new residual.
When the fractional difference in the standard deviation between iterations was less than 1\%, we stopped iterating and assigned the last standard deviation as $\Delta d$.
We found that $\Delta d=1~\mathrm{\mu m}$ for all sensor locations. 
This is larger than the nominal dynamic uncertainty (0.2~µm) due to the RHWP system's vibration during rotation.


\subsection{Wire--Mirror Separation}\label{ssec:WMS}

\begin{figure}
    \centering
    \includegraphics[width=\linewidth]{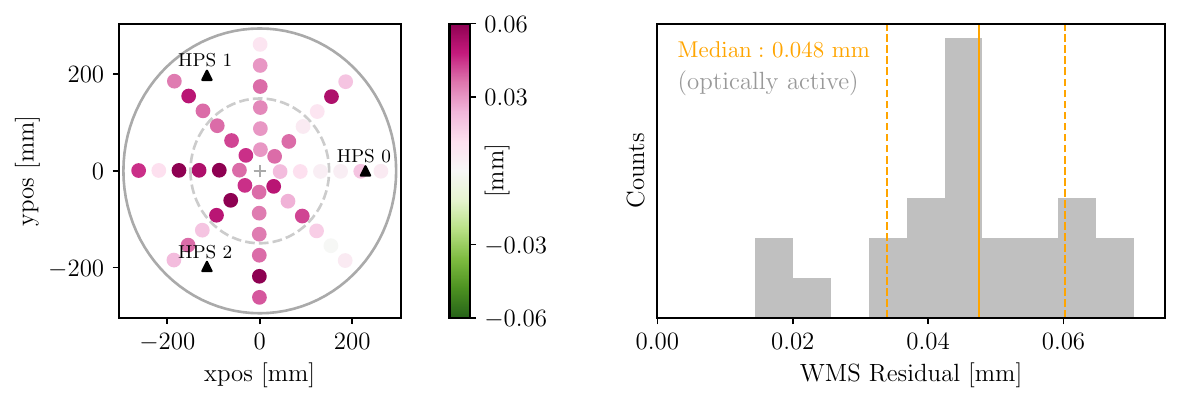}
    \caption{\textit{Left}: the WMS residual (relative to the ideal $z_0=0.88$~mm), measured with the microscope and camera. 
    The mirror maximum (optically active) region is delineated with a gray solid (dashed) line. The locations of the three HPSs where adjustments were made are marked with black triangles. 
    \textit{Right}: the histogram of the WMS residual measurements within the optically active region. The median and 16/84th quantile of the WMS residual is $0.048^{+0.013}_{-0.014}$~mm, which results in a modulation efficiency $\epsilon=96.2^{-0.4}_{+0.5}\%$ with the estimated bandpass.\cite{dahal22}}
    \label{fig:WMS}
\end{figure}

The wire--mirror separation (WMS, denoted with $z$ as in Figure \ref{fig:model}) directly affects the modulation efficiency. 

If the phase delay changes as $\widetilde\phi=\pi+\delta\phi$, the outgoing Stokes parameters are altered as:
\begin{equation}
\begin{aligned}
    \widetilde I^{\mathrm{out}}(t)&=I^{\mathrm{in}},\\
    \widetilde Q^{\mathrm{out}}(t)&=A~Q^{\mathrm{in}}
    +B~U^{\mathrm{in}}
    +\sin2\omega t\sin\delta\phi ~V^{\mathrm{in}}
    +\sin^2\frac{\delta\phi}2~Q^\mathrm{in},\\
    \widetilde U^{\mathrm{out}}(t)&=B~Q^{\mathrm{in}}
    -A~U^{\mathrm{in}}
    +\cos2\omega t\sin\delta\phi ~V^{\mathrm{in}}
    +\sin^2\frac{\delta\phi}2~U^\mathrm{in},\\
    \widetilde V^{\mathrm{out}}(t)&=\sin 2\omega t\sin\delta\phi~Q^{\mathrm{in}}
    -\cos 2\omega t\sin\delta\phi~U^{\mathrm{in}}
    -\cos\delta\phi~V^{\mathrm{in}},
\end{aligned}
\end{equation}
where
\begin{equation}
\begin{aligned}
    A=\cos4\omega t\cos^{2}\frac{\delta\phi}2
    ,\quad B=\sin4\omega t\cos^2\frac{\delta\phi}2.
\end{aligned}
\end{equation}
The monochromatic modulation efficiency is given by \cite{komatsu2020design, adler2024modeling}:
\begin{equation}
    \epsilon_\nu=\frac{\sqrt{(A+A)^2+(B+B)^2}}{2+A-A}=\cos^2\frac{\delta\phi_\nu}{2}.
\end{equation}
In reality, a nonzero $\delta\phi_\nu$ (the phase-delay deviation for a light at frequency $\nu$) can result from WMS misalignment and finite detector bandwidth as:
\begin{equation}
    \delta\phi_\nu=\frac{4\pi\cos\theta}{c}(z\nu-z_0\nu_0),
\end{equation}
where $\nu_{0}=92$~GHz is the bandcenter for the CLASS 90 GHz telescope\cite{dahal22}, and $z_0=0.88$~mm.
Finally, the modulation efficiency across the band is:
\begin{equation}
    \epsilon=\frac{\int\epsilon_{\nu}f(\nu)d\nu}{\int f(\nu)d\nu},
\end{equation}
where $f(\nu)$ is the bandpass.


We used the microscope to map the WMS by measuring the linear stage travel between focusing on the wires and the mirror.
We adjusted the WMS using the HPSs based on the measured results until satisfied.
The final characterization includes 48 measurements in total.
The left panel of Figure \ref{fig:WMS} maps the WMS residual, and the right panel shows the histogram of the measurements in the optically active region ($<30$~cm)\cite{eimer2022spie}.
The median and 16/84th percentile of the WMS residual are $0.048^{+0.013}_{-0.014}$~mm, resulting in an estimated modulation efficiency $\epsilon=96.2^{-0.4}_{+0.5}\%$ with an estimated bandpass \cite{dahal22} (top-hat profile, 34~GHz width, centered at 92~GHz).


\section{Rotation Stability}\label{sec:mechanical}
\begin{figure}[h]
    \centering
    \includegraphics[width=\linewidth]{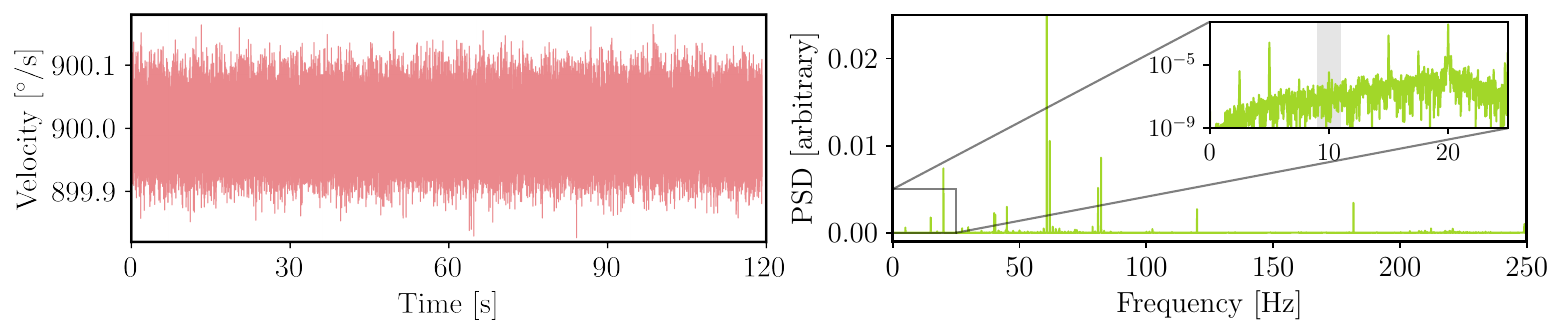}
    \caption{\textit{Left}: the RHWP angular velocity during a 2-min operation, and the RHWP maintained a constant angular velocity to an accuracy of within 0.005\%.
    \textit{Right}: the PSD of the RHWP angular velocity. No significant feature was found near the signal band (gray vertical band, $\sim$10~Hz).}
    \label{fig:stability}
\end{figure}

In-lab mechanical tests were performed to assess the rotation stability of the RHWP.
The mirror surface was positioned at $26^\circ$ from vertical, simulating the nominal orientation for CMB scans at a $45^\circ$ elevation and a $0^\circ$ boresight angle.
The motor was set to rotate at 2.5~Hz (900$^\circ$/s). Figure \ref{fig:stability} shows results from a 2-minute operation.
The RHWP angular velocity remained stable, with a fractional standard deviation of less than 0.005\%.
The power spectral density (PSD) of the angular velocity over the 2-minute rotation period showed no significant features near the signal band ($\sim$10~Hz).

~\\

\section{Conclusions}\label{sec:conclusion}
Polarization modulation is an effective method for achieving long-time instrumental stability, critical for measuring the large angular scales of the CMB polarization.
This work presents the design and in-lab performance test results of the first RHWP for the CLASS telescopes. 

The design of the support plate and the coupling of RHWP and the drive system has been simplified compared to E22 \cite{eimer2022spie}.
The mounting structure allows for adjustments in the orientation and the position of the RHWP system.
A one-dimensional profilometer was used to characterize the RHWP properties, and a custom-designed stretching frame was employed to tension the wires to the desired profile.

For the in-lab test results, the wire resonance frequency measurements closely match the target. 
The mirror tilt relative to the rotation axis was controlled to be within $15''$ at multiple radii, which corresponds to a negligible $0.6''$ beam smearing amplitude.
The median and 16/84th percentile of the wire--mirror separation residual was maintained at $0.048^{+0.013}_{-0.014}$~mm in the optically active region, which results in a modulation efficiency $\epsilon=96.2^{-0.4}_{+0.5}\%$ with an estimated bandpass of 34~GHz\cite{dahal22}.
Finally, the RHWP sustained a constant rotation frequency of 2.5~Hz to an accuracy of within 0.005\%.

The RHWP has been successfully integrated into the existing 90 GHz telescope and started collecting data in June 2024.
In addition to cosmological analyses, the data with the RHWP in operation will also be used to address sources of systematic errors, thereby enhancing the telescope's design and optimizing its performance to better achieve the scientific objectives.





\acknowledgments 
We acknowledge the National Science Foundation Division of Astronomical Sciences for their support. 
The RHWP modulator has been developed with support from Grant Numbers 2034400 and 2109311. We thank Hayley Nofi for helping build the RHWP wire array. We further acknowledge the very generous support of Jim Murren and Heather Miller (JHU A\&S ’88), Matthew Polk (JHU A\&S Physics BS ’71), David Nicholson, and Michael Bloomberg (JHU Engineering ’64). 

\bibliography{main, class_pub, cmb, cosmology, software, Planck_bib} 

\begin{thebibliography}{10}

\bibitem{hu1997cmb}
{Hu}, W. and {White}, M., ``{A CMB polarization primer},'' {\em \na}~{\bf 2},  323--344 (Oct. 1997).

\bibitem{mukhanov2005physical}
{Mukhanov}, V.,  [{\em {Physical Foundations of Cosmology}}{\nolinebreak\hspace{0.1em}]} (2005).

\bibitem{dodelson2020modern}
{Dodelson}, S. and {Schmidt}, F.,  [{\em {Modern Cosmology}}{\nolinebreak\hspace{0.1em}]} (2020).

\bibitem{zaldarriaga1997all}
{Zaldarriaga}, M. and {Seljak}, U., ``{All-sky analysis of polarization in the microwave background},'' {\em \prd}~{\bf 55},  1830--1840 (Feb. 1997).

\bibitem{kamionkowski1997statistics}
{Kamionkowski}, M., {Kosowsky}, A., and {Stebbins}, A., ``{Statistics of cosmic microwave background polarization},'' {\em \prd}~{\bf 55},  7368--7388 (June 1997).

\bibitem{baumann2009probing}
{Baumann}, D., {Jackson}, M.~G., {Adshead}, P., {Amblard}, A., {Ashoorioon}, A., {Bartolo}, N., {Bean}, R., {Beltr{\'a}n}, M., {de Bernardis}, F., {Bird}, S., {Chen}, X., {Chung}, D. J.~H., {Colombo}, L., {Cooray}, A., {Creminelli}, P., {Dodelson}, S., {Dunkley}, J., {Dvorkin}, C., {Easther}, R., {Finelli}, F., {Flauger}, R., {Hertzberg}, M.~P., {Jones-Smith}, K., {Kachru}, S., {Kadota}, K., {Khoury}, J., {Kinney}, W.~H., {Komatsu}, E., {Krauss}, L.~M., {Lesgourgues}, J., {Liddle}, A., {Liguori}, M., {Lim}, E., {Linde}, A., {Matarrese}, S., {Mathur}, H., {McAllister}, L., {Melchiorri}, A., {Nicolis}, A., {Pagano}, L., {Peiris}, H.~V., {Peloso}, M., {Pogosian}, L., {Pierpaoli}, E., {Riotto}, A., {Seljak}, U., {Senatore}, L., {Shandera}, S., {Silverstein}, E., {Smith}, T., {Vaudrevange}, P., {Verde}, L., {Wandelt}, B., {Wands}, D., {Watson}, S., {Wyman}, M., {Yadav}, A., {Valkenburg}, W., and {Zaldarriaga}, M., ``{Probing Inflation with CMB Polarization},'' in [{\em CMB Polarization Workshop: Theory and
  Foregrounds: CMBPol Mission Concept Study}{\nolinebreak\hspace{0.1em}]},  {Dodelson}, S., {Baumann}, D., {Cooray}, A., {Dunkley}, J., {Fraisse}, A., {Jackson}, M.~G., {Kogut}, A., {Krauss}, L., {Zaldarriaga}, M., and {Smith}, K., eds., {\em American Institute of Physics Conference Series} {\bf 1141},  10--120, AIP (June 2009).

\bibitem{abazajian2015inflation}
{Abazajian}, K.~N., {Arnold}, K., {Austermann}, J., {Benson}, B.~A., {Bischoff}, C., {Bock}, J., {Bond}, J.~R., {Borrill}, J., {Buder}, I., {Burke}, D.~L., {Calabrese}, E., {Carlstrom}, J.~E., {Carvalho}, C.~S., {Chang}, C.~L., {Chiang}, H.~C., {Church}, S., {Cooray}, A., {Crawford}, T.~M., {Crill}, B.~P., {Dawson}, K.~S., {Das}, S., {Devlin}, M.~J., {Dobbs}, M., {Dodelson}, S., {Dor{\'e}}, O., {Dunkley}, J., {Feng}, J.~L., {Fraisse}, A., {Gallicchio}, J., {Giddings}, S.~B., {Green}, D., {Halverson}, N.~W., {Hanany}, S., {Hanson}, D., {Hildebrandt}, S.~R., {Hincks}, A., {Hlozek}, R., {Holder}, G., {Holzapfel}, W.~L., {Honscheid}, K., {Horowitz}, G., {Hu}, W., {Hubmayr}, J., {Irwin}, K., {Jackson}, M., {Jones}, W.~C., {Kallosh}, R., {Kamionkowski}, M., {Keating}, B., {Keisler}, R., {Kinney}, W., {Knox}, L., {Komatsu}, E., {Kovac}, J., {Kuo}, C.~L., {Kusaka}, A., {Lawrence}, C., {Lee}, A.~T., {Leitch}, E., {Linde}, A., {Linder}, E., {Lubin}, P., {Maldacena}, J., {Martinec}, E., {McMahon}, J., {Miller}, A.,
  {Mukhanov}, V., {Newburgh}, L., {Niemack}, M.~D., {Nguyen}, H., {Nguyen}, H.~T., {Page}, L., {Pryke}, C., {Reichardt}, C.~L., {Ruhl}, J.~E., {Sehgal}, N., {Seljak}, U., {Senatore}, L., {Sievers}, J., {Silverstein}, E., {Slosar}, A., {Smith}, K.~M., {Spergel}, D., {Staggs}, S.~T., {Stark}, A., {Stompor}, R., {Vieregg}, A.~G., {Wang}, G., {Watson}, S., {Wollack}, E.~J., {Wu}, W.~L.~K., {Yoon}, K.~W., {Zahn}, O., and {Zaldarriaga}, M., ``{Inflation physics from the cosmic microwave background and large scale structure},'' {\em Astroparticle Physics}~{\bf 63},  55--65 (Mar. 2015).

\bibitem{bennett2012}
{Bennett}, C.~L., {Larson}, D., {Weiland}, J.~L., {Jarosik}, N., {Hinshaw}, G., {Odegard}, N., {Smith}, K.~M., {Hill}, R.~S., {Gold}, B., {Halpern}, M., {Komatsu}, E., {Nolta}, M.~R., {Page}, L., {Spergel}, D.~N., {Wollack}, E., {Dunkley}, J., {Kogut}, A., {Limon}, M., {Meyer}, S.~S., {Tucker}, G.~S., and {Wright}, E.~L., ``{Nine-year Wilkinson Microwave Anisotropy Probe (WMAP) Observations: Final Maps and Results},'' {\em \apjs}~{\bf 208},  20 (Oct. 2013).

\bibitem{hinshaw13}
{Hinshaw}, G., {Larson}, D., {Komatsu}, E., {Spergel}, D.~N., {Bennett}, C.~L., {Dunkley}, J., {Nolta}, M.~R., {Halpern}, M., {Hill}, R.~S., {Odegard}, N., {Page}, L., {Smith}, K.~M., {Weiland}, J.~L., {Gold}, B., {Jarosik}, N., {Kogut}, A., {Limon}, M., {Meyer}, S.~S., {Tucker}, G.~S., {Wollack}, E., and {Wright}, E.~L., ``{Nine-year Wilkinson Microwave Anisotropy Probe (WMAP) Observations: Cosmological Parameter Results},'' {\em \apjs}~{\bf 208},  19 (Oct. 2013).

\bibitem{planck18IV}
{Planck Collaboration}, {Akrami}, Y., {Ashdown}, M., {Aumont}, J., {Baccigalupi}, C., {Ballardini}, M., {Banday}, A.~J., {Barreiro}, R.~B., {Bartolo}, N., {Basak}, S., {Benabed}, K., {Bersanelli}, M., {Bielewicz}, P., {Bond}, J.~R., {Borrill}, J., {Bouchet}, F.~R., {Boulanger}, F., {Bucher}, M., {Burigana}, C., {Calabrese}, E., {Cardoso}, J.~F., {Carron}, J., {Casaponsa}, B., {Challinor}, A., {Colombo}, L.~P.~L., {Combet}, C., {Crill}, B.~P., {Cuttaia}, F., {de Bernardis}, P., {de Rosa}, A., {de Zotti}, G., {Delabrouille}, J., {Delouis}, J.~M., {Di Valentino}, E., {Dickinson}, C., {Diego}, J.~M., {Donzelli}, S., {Dor{\'e}}, O., {Ducout}, A., {Dupac}, X., {Efstathiou}, G., {Elsner}, F., {En{\ss}lin}, T.~A., {Eriksen}, H.~K., {Falgarone}, E., {Fernandez-Cobos}, R., {Finelli}, F., {Forastieri}, F., {Frailis}, M., {Fraisse}, A.~A., {Franceschi}, E., {Frolov}, A., {Galeotta}, S., {Galli}, S., {Ganga}, K., {G{\'e}nova-Santos}, R.~T., {Gerbino}, M., {Ghosh}, T., {Gonz{\'a}lez-Nuevo}, J., {G{\'o}rski}, K.~M.,
  {Gratton}, S., {Gruppuso}, A., {Gudmundsson}, J.~E., {Handley}, W., {Hansen}, F.~K., {Helou}, G., {Herranz}, D., {Hildebrandt}, S.~R., {Huang}, Z., {Jaffe}, A.~H., {Karakci}, A., {Keih{\"a}nen}, E., {Keskitalo}, R., {Kiiveri}, K., {Kim}, J., {Kisner}, T.~S., {Krachmalnicoff}, N., {Kunz}, M., {Kurki-Suonio}, H., {Lagache}, G., {Lamarre}, J.~M., {Lasenby}, A., {Lattanzi}, M., {Lawrence}, C.~R., {Le Jeune}, M., {Levrier}, F., {Liguori}, M., {Lilje}, P.~B., {Lindholm}, V., {L{\'o}pez-Caniego}, M., {Lubin}, P.~M., {Ma}, Y.~Z., {Mac{\'\i}as-P{\'e}rez}, J.~F., {Maggio}, G., {Maino}, D., {Mandolesi}, N., {Mangilli}, A., {Marcos-Caballero}, A., {Maris}, M., {Martin}, P.~G., {Mart{\'\i}nez-Gonz{\'a}lez}, E., {Matarrese}, S., {Mauri}, N., {McEwen}, J.~D., {Meinhold}, P.~R., {Melchiorri}, A., {Mennella}, A., {Migliaccio}, M., {Miville-Desch{\^e}nes}, M.~A., {Molinari}, D., {Moneti}, A., {Montier}, L., {Morgante}, G., {Natoli}, P., {Oppizzi}, F., {Pagano}, L., {Paoletti}, D., {Partridge}, B., {Peel}, M., {Pettorino},
  V., {Piacentini}, F., {Polenta}, G., {Puget}, J.~L., {Rachen}, J.~P., {Reinecke}, M., {Remazeilles}, M., {Renzi}, A., {Rocha}, G., {Roudier}, G., {Rubi{\~n}o-Mart{\'\i}n}, J.~A., {Ruiz-Granados}, B., {Salvati}, L., {Sandri}, M., {Savelainen}, M., {Scott}, D., {Seljebotn}, D.~S., {Sirignano}, C., {Spencer}, L.~D., {Suur-Uski}, A.~S., {Tauber}, J.~A., {Tavagnacco}, D., {Tenti}, M., {Thommesen}, H., {Toffolatti}, L., {Tomasi}, M., {Trombetti}, T., {Valiviita}, J., {Van Tent}, B., {Vielva}, P., {Villa}, F., {Vittorio}, N., {Wandelt}, B.~D., {Wehus}, I.~K., {Zacchei}, A., and {Zonca}, A., ``{Planck 2018 results. IV. Diffuse component separation},'' {\em \aap}~{\bf 641},  A4 (Sept. 2020).

\bibitem{planck18VI}
{Planck Collaboration}, {Aghanim}, N., {Akrami}, Y., {Ashdown}, M., {Aumont}, J., {Baccigalupi}, C., {Ballardini}, M., {Banday}, A.~J., {Barreiro}, R.~B., {Bartolo}, N., {Basak}, S., {Battye}, R., {Benabed}, K., {Bernard}, J.~P., {Bersanelli}, M., {Bielewicz}, P., {Bock}, J.~J., {Bond}, J.~R., {Borrill}, J., {Bouchet}, F.~R., {Boulanger}, F., {Bucher}, M., {Burigana}, C., {Butler}, R.~C., {Calabrese}, E., {Cardoso}, J.~F., {Carron}, J., {Challinor}, A., {Chiang}, H.~C., {Chluba}, J., {Colombo}, L.~P.~L., {Combet}, C., {Contreras}, D., {Crill}, B.~P., {Cuttaia}, F., {de Bernardis}, P., {de Zotti}, G., {Delabrouille}, J., {Delouis}, J.~M., {Di Valentino}, E., {Diego}, J.~M., {Dor{\'e}}, O., {Douspis}, M., {Ducout}, A., {Dupac}, X., {Dusini}, S., {Efstathiou}, G., {Elsner}, F., {En{\ss}lin}, T.~A., {Eriksen}, H.~K., {Fantaye}, Y., {Farhang}, M., {Fergusson}, J., {Fernandez-Cobos}, R., {Finelli}, F., {Forastieri}, F., {Frailis}, M., {Fraisse}, A.~A., {Franceschi}, E., {Frolov}, A., {Galeotta}, S., {Galli}, S.,
  {Ganga}, K., {G{\'e}nova-Santos}, R.~T., {Gerbino}, M., {Ghosh}, T., {Gonz{\'a}lez-Nuevo}, J., {G{\'o}rski}, K.~M., {Gratton}, S., {Gruppuso}, A., {Gudmundsson}, J.~E., {Hamann}, J., {Handley}, W., {Hansen}, F.~K., {Herranz}, D., {Hildebrandt}, S.~R., {Hivon}, E., {Huang}, Z., {Jaffe}, A.~H., {Jones}, W.~C., {Karakci}, A., {Keih{\"a}nen}, E., {Keskitalo}, R., {Kiiveri}, K., {Kim}, J., {Kisner}, T.~S., {Knox}, L., {Krachmalnicoff}, N., {Kunz}, M., {Kurki-Suonio}, H., {Lagache}, G., {Lamarre}, J.~M., {Lasenby}, A., {Lattanzi}, M., {Lawrence}, C.~R., {Le Jeune}, M., {Lemos}, P., {Lesgourgues}, J., {Levrier}, F., {Lewis}, A., {Liguori}, M., {Lilje}, P.~B., {Lilley}, M., {Lindholm}, V., {L{\'o}pez-Caniego}, M., {Lubin}, P.~M., {Ma}, Y.~Z., {Mac{\'\i}as-P{\'e}rez}, J.~F., {Maggio}, G., {Maino}, D., {Mandolesi}, N., {Mangilli}, A., {Marcos-Caballero}, A., {Maris}, M., {Martin}, P.~G., {Martinelli}, M., {Mart{\'\i}nez-Gonz{\'a}lez}, E., {Matarrese}, S., {Mauri}, N., {McEwen}, J.~D., {Meinhold}, P.~R., {Melchiorri},
  A., {Mennella}, A., {Migliaccio}, M., {Millea}, M., {Mitra}, S., {Miville-Desch{\^e}nes}, M.~A., {Molinari}, D., {Montier}, L., {Morgante}, G., {Moss}, A., {Natoli}, P., {N{\o}rgaard-Nielsen}, H.~U., {Pagano}, L., {Paoletti}, D., {Partridge}, B., {Patanchon}, G., {Peiris}, H.~V., {Perrotta}, F., {Pettorino}, V., {Piacentini}, F., {Polastri}, L., {Polenta}, G., {Puget}, J.~L., {Rachen}, J.~P., {Reinecke}, M., {Remazeilles}, M., {Renzi}, A., {Rocha}, G., {Rosset}, C., {Roudier}, G., {Rubi{\~n}o-Mart{\'\i}n}, J.~A., {Ruiz-Granados}, B., {Salvati}, L., {Sandri}, M., {Savelainen}, M., {Scott}, D., {Shellard}, E.~P.~S., {Sirignano}, C., {Sirri}, G., {Spencer}, L.~D., {Sunyaev}, R., {Suur-Uski}, A.~S., {Tauber}, J.~A., {Tavagnacco}, D., {Tenti}, M., {Toffolatti}, L., {Tomasi}, M., {Trombetti}, T., {Valenziano}, L., {Valiviita}, J., {Van Tent}, B., {Vibert}, L., {Vielva}, P., {Villa}, F., {Vittorio}, N., {Wandelt}, B.~D., {Wehus}, I.~K., {White}, M., {White}, S.~D.~M., {Zacchei}, A., and {Zonca}, A., ``{Planck 2018
  results. VI. Cosmological parameters},'' {\em \aap}~{\bf 641},  A6 (Sept. 2020).

\bibitem{kusaka2018results}
{Kusaka}, A., {Appel}, J., {Essinger-Hileman}, T., {Beall}, J.~A., {Campusano}, L.~E., {Cho}, H.-M., {Choi}, S.~K., {Crowley}, K., {Fowler}, J.~W., {Gallardo}, P., {Hasselfield}, M., {Hilton}, G., {Ho}, S.-P.~P., {Irwin}, K., {Jarosik}, N., {Niemack}, M.~D., {Nixon}, G.~W., {\raisebox{-0.5ex}\textasciitilde Nolta}, M., {Page}, Lyman~A., J., {Palma}, G.~A., {Parker}, L., {Raghunathan}, S., {Reintsema}, C.~D., {Sievers}, J., {Simon}, S.~M., {Staggs}, S.~T., {Visnjic}, K., and {Yoon}, K.-W., ``{Results from the Atacama B-mode Search (ABS) experiment},'' {\em \jcap}~{\bf 2018},  005 (Sept. 2018).

\bibitem{BK-XVII23}
{BICEP/Keck Collaboration}, {Ade}, P.~A.~R., {Ahmed}, Z., {Amiri}, M., {Barkats}, D., {Thakur}, R.~B., {Bischoff}, C.~A., {Beck}, D., {Bock}, J.~J., {Boenish}, H., {Bullock}, E., {Buza}, V., {Cheshire}, J.~R., I., {Connors}, J., {Cornelison}, J., {Crumrine}, M., {Cukierman}, A., {Denison}, E.~V., {Dierickx}, M., {Duband}, L., {Eiben}, M., {Fatigoni}, S., {Filippini}, J.~P., {Fliescher}, S., {Giannakopoulos}, C., {Goeckner-Wald}, N., {Goldfinger}, D.~C., {Grayson}, J., {Grimes}, P., {Hall}, G., {Halal}, G., {Halpern}, M., {Hand}, E., {Harrison}, S., {Henderson}, S., {Hildebrandt}, S.~R., {Hubmayr}, J., {Hui}, H., {Irwin}, K.~D., {Kang}, J., {Karkare}, K.~S., {Karpel}, E., {Kefeli}, S., {Kernasovskiy}, S.~A., {Kovac}, J.~M., {Kuo}, C.~L., {Lau}, K., {Leitch}, E.~M., {Lennox}, A., {Megerian}, K.~G., {Minutolo}, L., {Moncelsi}, L., {Nakato}, Y., {Namikawa}, T., {Nguyen}, H.~T., {O'Brient}, R., {Ogburn}, R.~W., I., {Palladino}, S., {Petroff}, M., {Prouve}, T., {Pryke}, C., {Racine}, B., {Reintsema}, C.~D.,
  {Richter}, S., {Schillaci}, A., {Schwarz}, R., {Schmitt}, B.~L., {Sheehy}, C.~D., {Singari}, B., {Soliman}, A., {St. Germaine}, T., {Steinbach}, B., {Sudiwala}, R.~V., {Teply}, G.~P., {Thompson}, K.~L., {Tolan}, J.~E., {Tucker}, C., {Turner}, A.~D., {Umilt{\`a}}, C., {Verg{\`e}s}, C., {Vieregg}, A.~G., {Wandui}, A., {Weber}, A.~C., {Wiebe}, D.~V., {Willmert}, J., {Wong}, C.~L., {Wu}, W.~L.~K., {Yang}, H., {Yoon}, K.~W., {Young}, E., {Yu}, C., {Zeng}, L., {Zhang}, C., {Zhang}, S., and {Bicep/Keck Collaboration}, ``{BICEP/Keck. XVII. Line-of-sight Distortion Analysis: Estimates of Gravitational Lensing, Anisotropic Cosmic Birefringence, Patchy Reionization, and Systematic Errors},'' {\em \apj}~{\bf 949},  43 (June 2023).

\bibitem{eimer23}
{Eimer}, J.~R., {Li}, Y., {Brewer}, M.~K., {Shi}, R., {Ali}, A., {Appel}, J.~W., {Bennett}, C.~L., {Bruno}, S.~M., {Bustos}, R., {Chuss}, D.~T., {Cleary}, J., {Dahal}, S., {Datta}, R., {Denes Couto}, J., {Denis}, K.~L., {D{\"u}nner}, R., {Essinger-Hileman}, T., {Flux{\'a}}, P., {Hubmayer}, J., {Harrington}, K., {Iuliano}, J., {Karakla}, J., {Marriage}, T.~A., {N{\'u}{\~n}ez}, C., {Parker}, L., {Petroff}, M.~A., {Reeves}, R.~A., {Rostem}, K., {Valle}, D. A.~N., {Watts}, D.~J., {Weiland}, J.~L., {Wollack}, E.~J., {Xu}, Z., and {Zeng}, L., ``{CLASS Angular Power Spectra and Map-component Analysis for 40 GHz Observations through 2022},'' {\em \apj}~{\bf 963},  92 (Mar. 2024).

\bibitem{adachi2022improved}
{Adachi}, S., {Adkins}, T., {Aguilar Fa{\'u}ndez}, M.~A.~O., {Arnold}, K.~S., {Baccigalupi}, C., {Barron}, D., {Chapman}, S., {Cheung}, K., {Chinone}, Y., {Crowley}, K.~T., {Elleflot}, T., {Errard}, J., {Fabbian}, G., {Feng}, C., {Fujino}, T., {Galitzki}, N., {Halverson}, N.~W., {Hasegawa}, M., {Hazumi}, M., {Hirose}, H., {Howe}, L., {Ito}, J., {Jeong}, O., {Kaneko}, D., {Katayama}, N., {Keating}, B., {Kisner}, T., {Krachmalnicoff}, N., {Kusaka}, A., {Lee}, A.~T., {Linder}, E., {Lonappan}, A.~I., {Lowry}, L.~N., {Matsuda}, F., {Matsumura}, T., {Minami}, Y., {Murata}, M., {Nishino}, H., {Nishinomiya}, Y., {Poletti}, D., {Reichardt}, C.~L., {Ross}, C., {Segawa}, Y., {Siritanasak}, P., {Stompor}, R., {Suzuki}, A., {Tajima}, O., {Takakura}, S., {Takatori}, S., {Tanabe}, D., {Teply}, G., {Yamada}, K., {Zhou}, Y., and {POLARBEAR Collaboration}, ``{Improved Upper Limit on Degree-scale CMB B-mode Polarization Power from the 670 Square-degree POLARBEAR Survey},'' {\em \apj}~{\bf 931},  101 (June 2022).

\bibitem{quijoteIV}
{Rubi{\~n}o-Mart{\'\i}n}, J.~A., {Guidi}, F., {G{\'e}nova-Santos}, R.~T., {Harper}, S.~E., {Herranz}, D., {Hoyland}, R.~J., {Lasenby}, A.~N., {Poidevin}, F., {Rebolo}, R., {Ruiz-Granados}, B., {Vansyngel}, F., {Vielva}, P., {Watson}, R.~A., {Artal}, E., {Ashdown}, M., {Barreiro}, R.~B., {Bilbao-Ahedo}, J.~D., {Casas}, F.~J., {Casaponsa}, B., {Cepeda-Arroita}, R., {de la Hoz}, E., {Dickinson}, C., {Fern{\'a}ndez-Cobos}, R., {Fern{\'a}ndez-Torreiro}, M., {Gonz{\'a}lez-Gonz{\'a}lez}, R., {Hern{\'a}ndez-Monteagudo}, C., {L{\'o}pez-Caniego}, M., {L{\'o}pez-Caraballo}, C., {Mart{\'\i}nez-Gonz{\'a}lez}, E., {Peel}, M.~W., {Pel{\'a}ez-Santos}, A.~E., {Perrott}, Y., {Piccirillo}, L., {Razavi-Ghods}, N., {Scott}, P., {Titterington}, D., {Tramonte}, D., and {Vignaga}, R., ``{QUIJOTE scientific results - IV. A northern sky survey in intensity and polarization at 10-20 GHz with the multifrequency instrument},'' {\em \mnras}~{\bf 519},  3383--3431 (Mar. 2023).

\bibitem{ansarinejad2024mass}
{Ansarinejad}, B., {Raghunathan}, S., {Abbott}, T.~M.~C., {Ade}, P.~A.~R., {Aguena}, M., {Alves}, O., {Anderson}, A.~J., {Andrade-Oliveira}, F., {Archipley}, M., {Balkenhol}, L., {Benabed}, K., {Bender}, A.~N., {Benson}, B.~A., {Bertin}, E., {Bianchini}, F., {Bleem}, L.~E., {Bocquet}, S., {Bouchet}, F.~R., {Brooks}, D., {Bryant}, L., {Burke}, D.~L., {Camphuis}, E., {Carlstrom}, J.~E., {Carnero Rosell}, A., {Carretero}, J., {Castander}, F.~J., {Cecil}, T.~W., {Chang}, C.~L., {Chaubal}, P., {Chichura}, P.~M., {Chou}, T.~L., {Coerver}, A., {Costanzi}, M., {Crawford}, T.~M., {Cukierman}, A., {da Costa}, L.~N., {Daley}, C., {Davis}, T.~M., {de Haan}, T., {Desai}, S., {De Vicente}, J., {Dibert}, K.~R., {Dobbs}, M.~A., {Doel}, P., {Doussot}, A., {Doux}, C., {Dutcher}, D., {Everett}, W., {Feng}, C., {Ferguson}, K.~R., {Ferrero}, I., {Fichman}, K., {Foster}, A., {Frieman}, J., {Galli}, S., {Gambrel}, A.~E., {Garc{\'\i}a-Bellido}, J., {Gardner}, R.~W., {Gaztanaga}, E., {Ge}, F., {Giannini}, G., {Goeckner-Wald}, N.,
  {Grandis}, S., {Gruendl}, R.~A., {Gualtieri}, R., {Guidi}, F., {Guns}, S., {Gutierrez}, G., {Halverson}, N.~W., {Hinton}, S.~R., {Hivon}, E., {Holder}, G.~P., {Hollowood}, D.~L., {Holzapfel}, W.~L., {Honscheid}, K., {Hood}, J.~C., {Huang}, N., {James}, D.~J., {K{\'e}ruzor{\'e}}, F., {Knox}, L., {Korman}, M., {Kuo}, C.~L., {Lee}, A.~T., {Lee}, S., {Levy}, K., {Lowitz}, A.~E., {Lu}, C., {Maniyar}, A., {Marshall}, J.~L., {Mena-Fern{\'a}ndez}, J., {Menanteau}, F., {Miquel}, R., {Millea}, M., {Mohr}, J.~J., {Montgomery}, J., {Nakato}, Y., {Natoli}, T., {Noble}, G.~I., {Novosad}, V., {Ogando}, R.~L.~C., {Omori}, Y., {Padin}, S., {Palmese}, A., {Pan}, Z., {Paschos}, P., {Pereira}, M.~E.~S., {Pieres}, A., {Plazas Malag{\'o}n}, A.~A., {Prabhu}, K., {Quan}, W., {Rahlin}, A., {Rahimi}, M., {Reichardt}, C.~L., {Reil}, K., {Romer}, A.~K., {Rouble}, M., {Ruhl}, J.~E., {Sanchez}, E., {Sanchez Cid}, D., {Schiappucci}, E., {Sevilla-Noarbe}, I., {Smecher}, G., {Smith}, M., {Sobrin}, J.~A., {Stark}, A.~A., {Stephen}, J.,
  {Suchyta}, E., {Suzuki}, A., {Swanson}, M.~E.~C., {Tandoi}, C., {Tarle}, G., {Thompson}, K.~L., {Thorne}, B., {Trendafilova}, C., {Tucker}, C., {Umilta}, C., {Vieira}, J.~D., {Wang}, G., {Weaverdyck}, N., {Whitehorn}, N., {Wiseman.}, P., {Wu}, W.~L.~K., {Yefremenko}, V., {Young}, M.~R., and {Zebrowski}, J.~A., ``{Mass calibration of DES Year-3 clusters via SPT-3G CMB cluster lensing},'' {\em arXiv e-prints} ,  arXiv:2404.02153 (Apr. 2024).

\bibitem{pisano2014development}
{Pisano}, G., {Maffei}, B., {Ng}, M.~W., {Haynes}, V., {Brown}, M., {Noviello}, F., {de Bernardis}, P., {Masi}, S., {Piacentini}, F., {Pagano}, L., {Salatino}, M., {Ellison}, B., {Henry}, M., {de Maagt}, P., and {Shortt}, B., ``{Development of large radii half-wave plates for CMB satellite missions},'' in [{\em Millimeter, Submillimeter, and Far-Infrared Detectors and Instrumentation for Astronomy VII}{\nolinebreak\hspace{0.1em}]},  {Holland}, W.~S. and {Zmuidzinas}, J., eds., {\em \procspie}~{\bf 9153},  915317 (July 2014).

\bibitem{chus12vpm}
{Chuss}, D.~T., {Wollack}, E.~J., {Henry}, R., {Hui}, H., {Juarez}, A.~J., {Krejny}, M., {Moseley}, S.~H., and {Novak}, G., ``{Properties of a variable-delay polarization modulator},'' {\em \ao}~{\bf 51},  197 (Jan. 2012).

\bibitem{essinger-hileman2011probing}
{Essinger-Hileman}, T., {\em {Probing Inflationary Cosmology: The Atacama B-Mode Search (ABS)}}, PhD thesis, Princeton University, New Jersey (Jan. 2011).

\bibitem{fissel2010balloon}
{Fissel}, L.~M., {Ade}, P. A.~R., {Angil{\`e}}, F.~E., {Benton}, S.~J., {Chapin}, E.~L., {Devlin}, M.~J., {Gandilo}, N.~N., {Gundersen}, J.~O., {Hargrave}, P.~C., {Hughes}, D.~H., {Klein}, J., {Korotkov}, A.~L., {Marsden}, G., {Matthews}, T.~G., {Moncelsi}, L., {Mroczkowski}, T.~K., {Netterfield}, C.~B., {Novak}, G., {Olmi}, L., {Pascale}, E., {Savini}, G., {Scott}, D., {Shariff}, J.~A., {Soler}, J.~D., {Thomas}, N.~E., {Truch}, M. D.~P., {Tucker}, C.~E., {Tucker}, G.~S., {Ward-Thompson}, D., and {Wiebe}, D.~V., ``{The balloon-borne large-aperture submillimeter telescope for polarimetry: BLAST-Pol},'' in [{\em Millimeter, Submillimeter, and Far-Infrared Detectors and Instrumentation for Astronomy V}{\nolinebreak\hspace{0.1em}]},  {Holland}, W.~S. and {Zmuidzinas}, J., eds., {\em \procspie}~{\bf 7741},  77410E (July 2010).

\bibitem{ebex2018}
{EBEX Collaboration}, {Aboobaker}, A.~M., {Ade}, P., {Araujo}, D., {Aubin}, F., {Baccigalupi}, C., {Bao}, C., {Chapman}, D., {Didier}, J., {Dobbs}, M., {Geach}, C., {Grainger}, W., {Hanany}, S., {Helson}, K., {Hillbrand}, S., {Hubmayr}, J., {Jaffe}, A., {Johnson}, B., {Jones}, T., {Klein}, J., {Korotkov}, A., {Lee}, A., {Levinson}, L., {Limon}, M., {MacDermid}, K., {Matsumura}, T., {Miller}, A.~D., {Milligan}, M., {Raach}, K., {Reichborn-Kjennerud}, B., {Sagiv}, I., {Savini}, G., {Spencer}, L., {Tucker}, C., {Tucker}, G.~S., {Westbrook}, B., {Young}, K., and {Zilic}, K., ``{The EBEX Balloon-borne Experiment{\textemdash}Optics, Receiver, and Polarimetry},'' {\em \apjs}~{\bf 239},  7 (Nov. 2018).

\bibitem{de2012swipe}
{de Bernardis}, P., {Aiola}, S., {Amico}, G., {Battistelli}, E., {Coppolecchia}, A., {Cruciani}, A., {D'Addabbo}, A., {D'Alessandro}, G., {De Gregori}, S., {De Petris}, M., {Goldie}, D., {Gualtieri}, R., {Haynes}, V., {Lamagna}, L., {Maffei}, B., {Masi}, S., {Nati}, F., {Ng}, M.~W., {Pagano}, L., {Piacentini}, F., {Piccirillo}, L., {Pisano}, G., {Romeo}, G., {Salatino}, M., {Schillaci}, A., {Tommasi}, E., and {Withington}, S., ``{SWIPE: a bolometric polarimeter for the Large-Scale Polarization Explorer},'' in [{\em Millimeter, Submillimeter, and Far-Infrared Detectors and Instrumentation for Astronomy VI}{\nolinebreak\hspace{0.1em}]},  {Holland}, W.~S. and {Zmuidzinas}, J., eds., {\em \procspie}~{\bf 8452},  84523F (Sept. 2012).

\bibitem{hazumi2019litebird}
{Hazumi}, M., {Ade}, P.~A.~R., {Akiba}, Y., {Alonso}, D., {Arnold}, K., {Aumont}, J., {Baccigalupi}, C., {Barron}, D., {Basak}, S., {Beckman}, S., {Borrill}, J., {Boulanger}, F., {Bucher}, M., {Calabrese}, E., {Chinone}, Y., {Cho}, S., {Cukierman}, A., {Curtis}, D.~W., {de Haan}, T., {Dobbs}, M., {Dominjon}, A., {Dotani}, T., {Duband}, L., {Ducout}, A., {Dunkley}, J., {Duval}, J.~M., {Elleflot}, T., {Eriksen}, H.~K., {Errard}, J., {Fischer}, J., {Fujino}, T., {Funaki}, T., {Fuskeland}, U., {Ganga}, K., {Goeckner-Wald}, N., {Grain}, J., {Halverson}, N.~W., {Hamada}, T., {Hasebe}, T., {Hasegawa}, M., {Hattori}, K., {Hattori}, M., {Hayes}, L., {Hidehira}, N., {Hill}, C.~A., {Hilton}, G., {Hubmayr}, J., {Ichiki}, K., {Iida}, T., {Imada}, H., {Inoue}, M., {Inoue}, Y., {Irwin}, K.~D., {Ishino}, H., {Jeong}, O., {Kanai}, H., {Kaneko}, D., {Kashima}, S., {Katayama}, N., {Kawasaki}, T., {Kernasovskiy}, S.~A., {Keskitalo}, R., {Kibayashi}, A., {Kida}, Y., {Kimura}, K., {Kisner}, T., {Kohri}, K., {Komatsu}, E.,
  {Komatsu}, K., {Kuo}, C.~L., {Kurinsky}, N.~A., {Kusaka}, A., {Lazarian}, A., {Lee}, A.~T., {Li}, D., {Linder}, E., {Maffei}, B., {Mangilli}, A., {Maki}, M., {Matsumura}, T., {Matsuura}, S., {Meilhan}, D., {Mima}, S., {Minami}, Y., {Mitsuda}, K., {Montier}, L., {Nagai}, M., {Nagasaki}, T., {Nagata}, R., {Nakajima}, M., {Nakamura}, S., {Namikawa}, T., {Naruse}, M., {Nishino}, H., {Nitta}, T., {Noguchi}, T., {Ogawa}, H., {Oguri}, S., {Okada}, N., {Okamoto}, A., {Okamura}, T., {Otani}, C., {Patanchon}, G., {Pisano}, G., {Rebeiz}, G., {Remazeilles}, M., {Richards}, P.~L., {Sakai}, S., {Sakurai}, Y., {Sato}, Y., {Sato}, N., {Sawada}, M., {Segawa}, Y., {Sekimoto}, Y., {Seljak}, U., {Sherwin}, B.~D., {Shimizu}, T., {Shinozaki}, K., {Stompor}, R., {Sugai}, H., {Sugita}, H., {Suzuki}, A., {Suzuki}, J., {Tajima}, O., {Takada}, S., {Takaku}, R., {Takakura}, S., {Takatori}, S., {Tanabe}, D., {Taylor}, E., {Thompson}, K.~L., {Thorne}, B., {Tomaru}, T., {Tomida}, T., {Tomita}, N., {Tristram}, M., {Tucker}, C., {Turin},
  P., {Tsujimoto}, M., {Uozumi}, S., {Utsunomiya}, S., {Uzawa}, Y., {Vansyngel}, F., {Wehus}, I.~K., {Westbrook}, B., {Willer}, M., {Whitehorn}, N., {Yamada}, Y., {Yamamoto}, R., {Yamasaki}, N., {Yamashita}, T., and {Yoshida}, M., ``{LiteBIRD: A Satellite for the Studies of B-Mode Polarization and Inflation from Cosmic Background Radiation Detection},'' {\em Journal of Low Temperature Physics}~{\bf 194},  443--452 (Mar. 2019).

\bibitem{johnson2007maxipol}
{Johnson}, B.~R., {Collins}, J., {Abroe}, M.~E., {Ade}, P.~A.~R., {Bock}, J., {Borrill}, J., {Boscaleri}, A., {de Bernardis}, P., {Hanany}, S., {Jaffe}, A.~H., {Jones}, T., {Lee}, A.~T., {Levinson}, L., {Matsumura}, T., {Rabii}, B., {Renbarger}, T., {Richards}, P.~L., {Smoot}, G.~F., {Stompor}, R., {Tran}, H.~T., {Winant}, C.~D., {Wu}, J.~H.~P., and {Zuntz}, J., ``{MAXIPOL: Cosmic Microwave Background Polarimetry Using a Rotating Half-Wave Plate},'' {\em \apj}~{\bf 665},  42--54 (Aug. 2007).

\bibitem{hill2016design}
{Hill}, C.~A., {Beckman}, S., {Chinone}, Y., {Goeckner-Wald}, N., {Hazumi}, M., {Keating}, B., {Kusaka}, A., {Lee}, A.~T., {Matsuda}, F., {Plambeck}, R., {Suzuki}, A., and {Takakura}, S., ``{Design and development of an ambient-temperature continuously-rotating achromatic half-wave plate for CMB polarization modulation on the POLARBEAR-2 experiment},'' in [{\em Millimeter, Submillimeter, and Far-Infrared Detectors and Instrumentation for Astronomy VIII}{\nolinebreak\hspace{0.1em}]},  {Holland}, W.~S. and {Zmuidzinas}, J., eds., {\em \procspie}~{\bf 9914},  99142U (July 2016).

\bibitem{wiesemeyer2014submillimeter}
{Wiesemeyer}, H., {Hezareh}, T., {Kreysa}, E., {Weiss}, A., {G{\"u}sten}, R., {Menten}, K.~M., {Siringo}, G., {Schuller}, F., and {Kovacs}, A., ``{Submillimeter Polarimetry with PolKa, a Reflection-Type Modulator for the APEX Telescope},'' {\em \pasp}~{\bf 126},  1027 (Nov. 2014).

\bibitem{yamada2024simons}
{Yamada}, K., {Bixler}, B., {Sakurai}, Y., {Ashton}, P.~C., {Sugiyama}, J., {Arnold}, K., {Begin}, J., {Corbett}, L., {Day-Weiss}, S., {Galitzki}, N., {Hill}, C.~A., {Johnson}, B.~R., {Jost}, B., {Kusaka}, A., {Koopman}, B.~J., {Lashner}, J., {Lee}, A.~T., {Mangu}, A., {Nishino}, H., {Page}, L.~A., {Randall}, M.~J., {Sasaki}, D., {Song}, X., {Spisak}, J., {Tsan}, T., {Wang}, Y., and {Williams}, P.~A., ``{The Simons Observatory: Cryogenic half wave plate rotation mechanism for the small aperture telescopes},'' {\em Review of Scientific Instruments}~{\bf 95},  024504 (Feb. 2024).

\bibitem{filippini2010spider}
{Filippini}, J.~P., {Ade}, P.~A.~R., {Amiri}, M., {Benton}, S.~J., {Bihary}, R., {Bock}, J.~J., {Bond}, J.~R., {Bonetti}, J.~A., {Bryan}, S.~A., {Burger}, B., {Chiang}, H.~C., {Contaldi}, C.~R., {Crill}, B.~P., {Dor{\'e}}, O., {Farhang}, M., {Fissel}, L.~M., {Gandilo}, N.~N., {Golwala}, S.~R., {Gudmundsson}, J.~E., {Halpern}, M., {Hasselfield}, M., {Hilton}, G., {Holmes}, W., {Hristov}, V.~V., {Irwin}, K.~D., {Jones}, W.~C., {Kuo}, C.~L., {MacTavish}, C.~J., {Mason}, P.~V., {Montroy}, T.~E., {Morford}, T.~A., {Netterfield}, C.~B., {O'Dea}, D.~T., {Rahlin}, A.~S., {Reintsema}, C.~D., {Ruhl}, J.~E., {Runyan}, M.~C., {Schenker}, M.~A., {Shariff}, J.~A., {Soler}, J.~D., {Trangsrud}, A., {Tucker}, C., {Tucker}, R.~S., and {Turner}, A.~D., ``{SPIDER: a balloon-borne CMB polarimeter for large angular scales},'' in [{\em Millimeter, Submillimeter, and Far-Infrared Detectors and Instrumentation for Astronomy V}{\nolinebreak\hspace{0.1em}]},  {Holland}, W.~S. and {Zmuidzinas}, J., eds., {\em \procspie}~{\bf 7741},
  77411N (July 2010).

\bibitem{krejny2008hertz}
{Krejny}, M., {Chuss}, D., {D'Aubigny}, C.~D., {Golish}, D., {Houde}, M., {Hui}, H., {Kulesa}, C., {Loewenstein}, R.~F., {Moseley}, S.~H., {Novak}, G., {Voellmer}, G., {Walker}, C., and {Wollack}, E., ``{The Hertz/VPM polarimeter: design and first light observations},'' {\em \ao}~{\bf 47},  4429 (Aug. 2008).

\bibitem{lazear2014primordial}
{Lazear}, J., {Ade}, P. A.~R., {Benford}, D., {Bennett}, C.~L., {Chuss}, D.~T., {Dotson}, J.~L., {Eimer}, J.~R., {Fixsen}, D.~J., {Halpern}, M., {Hilton}, G., {Hinderks}, J., {Hinshaw}, G.~F., {Irwin}, K., {Jhabvala}, C., {Johnson}, B., {Kogut}, A., {Lowe}, L., {McMahon}, J.~J., {Miller}, T.~M., {Mirel}, P., {Moseley}, S.~H., {Rodriguez}, S., {Sharp}, E., {Staguhn}, J.~G., {Switzer}, E.~R., {Tucker}, C.~E., {Weston}, A., and {Wollack}, E.~J., ``{The Primordial Inflation Polarization Explorer (PIPER)},'' in [{\em Millimeter, Submillimeter, and Far-Infrared Detectors and Instrumentation for Astronomy VII}{\nolinebreak\hspace{0.1em}]},  {Holland}, W.~S. and {Zmuidzinas}, J., eds., {\em \procspie}~{\bf 9153},  91531L (July 2014).

\bibitem{essinger-hileman14spie}
{Essinger-Hileman}, T., {Ali}, A., {Amiri}, M., {Appel}, J.~W., {Araujo}, D., {Bennett}, C.~L., {Boone}, F., {Chan}, M., {Cho}, H.-M., {Chuss}, D.~T., {Colazo}, F., {Crowe}, E., {Denis}, K., {D{\"u}nner}, R., {Eimer}, J., {Gothe}, D., {Halpern}, M., {Harrington}, K., {Hilton}, G.~C., {Hinshaw}, G.~F., {Huang}, C., {Irwin}, K., {Jones}, G., {Karakla}, J., {Kogut}, A.~J., {Larson}, D., {Limon}, M., {Lowry}, L., {Marriage}, T., {Mehrle}, N., {Miller}, A.~D., {Miller}, N., {Moseley}, S.~H., {Novak}, G., {Reintsema}, C., {Rostem}, K., {Stevenson}, T., {Towner}, D., {U-Yen}, K., {Wagner}, E., {Watts}, D., {Wollack}, E.~J., {Xu}, Z., and {Zeng}, L., ``{CLASS: the cosmology large angular scale surveyor},'' in [{\em Millimeter, Submillimeter, and Far-Infrared Detectors and Instrumentation for Astronomy VII}{\nolinebreak\hspace{0.1em}]},  {Holland}, W.~S. and {Zmuidzinas}, J., eds., {\em Society of Photo-Optical Instrumentation Engineers (SPIE) Conference Series} {\bf 9153},  91531I (July 2014).

\bibitem{harrington16spie}
{Harrington}, K., {Marriage}, T., {Ali}, A., {Appel}, J.~W., {Bennett}, C.~L., {Boone}, F., {Brewer}, M., {Chan}, M., {Chuss}, D.~T., {Colazo}, F., {Dahal}, S., {Denis}, K., {D{\"u}nner}, R., {Eimer}, J., {Essinger-Hileman}, T., {Fluxa}, P., {Halpern}, M., {Hilton}, G., {Hinshaw}, G.~F., {Hubmayr}, J., {Iuliano}, J., {Karakla}, J., {McMahon}, J., {Miller}, N.~T., {Moseley}, S.~H., {Palma}, G., {Parker}, L., {Petroff}, M., {Pradenas}, B., {Rostem}, K., {Sagliocca}, M., {Valle}, D., {Watts}, D., {Wollack}, E., {Xu}, Z., and {Zeng}, L., ``{The Cosmology Large Angular Scale Surveyor},'' in [{\em Millimeter, Submillimeter, and Far-Infrared Detectors and Instrumentation for Astronomy VIII}{\nolinebreak\hspace{0.1em}]},  {Holland}, W.~S. and {Zmuidzinas}, J., eds., {\em Society of Photo-Optical Instrumentation Engineers (SPIE) Conference Series} {\bf 9914},  99141K (July 2016).

\bibitem{eimer12spie}
{Eimer}, J.~R., {Bennett}, C.~L., {Chuss}, D.~T., {Marriage}, T., {Wollack}, E.~J., and {Zeng}, L., ``{The cosmology large angular scale surveyor (CLASS): 40 GHz optical design},'' in [{\em Millimeter, Submillimeter, and Far-Infrared Detectors and Instrumentation for Astronomy VI}{\nolinebreak\hspace{0.1em}]},  {Holland}, W.~S. and {Zmuidzinas}, J., eds., {\em Society of Photo-Optical Instrumentation Engineers (SPIE) Conference Series} {\bf 8452},  845220 (Sept. 2012).

\bibitem{harrington21}
{Harrington}, K., {Datta}, R., {Osumi}, K., {Ali}, A., {Appel}, J.~W., {Bennett}, C.~L., {Brewer}, M.~K., {Bustos}, R., {Chan}, M., {Chuss}, D.~T., {Cleary}, J., {Denes Couto}, J., {Dahal}, S., {D{\"u}nner}, R., {Eimer}, J.~R., {Essinger-Hileman}, T., {Hubmayr}, J., {Raul Espinoza Inostroza}, F., {Iuliano}, J., {Karakla}, J., {Li}, Y., {Marriage}, T.~A., {Miller}, N.~J., {N{\'u}{\~n}ez}, C., {Padilla}, I.~L., {Parker}, L., {Petroff}, M.~A., {Pradenas M{\'a}rquez}, B., {Reeves}, R., {Flux{\'a} Rojas}, P., {Rostem}, K., {Augusto Nunes Valle}, D., {Watts}, D.~J., {Weiland}, J.~L., {Wollack}, E.~J., {Xu}, Z., and {Class Collaboration}, ``{Two Year Cosmology Large Angular Scale Surveyor (CLASS) Observations: Long Timescale Stability Achieved with a Front-end Variable-delay Polarization Modulator at 40 GHz},'' {\em \apj}~{\bf 922},  212 (Dec. 2021).

\bibitem{cleary22}
{Cleary}, J., {Datta}, R., {Appel}, J.~W., {Bennet}, C.~L., {Chuss}, D.~T., {Denes Couto}, J., {Dahal}, S., {Espinoza}, F., {Essinger-Hileman}, T., {Harrington}, K., {Iuliano}, J., {Li}, Y., {Marriage}, T.~A., {N{\'u}{\~n}ez}, C., {Petroff}, M.~A., {Reeves}, R.~A., {Shi}, R., {Watts}, D.~J., {Wollack}, E.~J., and {Xu}, Z., ``{Long-timescale stability in CMB observations at multiple frequencies using front-end polarization modulation},'' in [{\em Millimeter, Submillimeter, and Far-Infrared Detectors and Instrumentation for Astronomy XI}{\nolinebreak\hspace{0.1em}]},  {Zmuidzinas}, J. and {Gao}, J.-R., eds., {\em Society of Photo-Optical Instrumentation Engineers (SPIE) Conference Series} {\bf 12190},  121902Q (Aug. 2022).

\bibitem{Li23}
{Li}, Y., {Eimer}, J.~R., {Osumi}, K., {Appel}, J.~W., {Brewer}, M.~K., {Ali}, A., {Bennett}, C.~L., {Bruno}, S.~M., {Bustos}, R., {Chuss}, D.~T., {Cleary}, J., {Couto}, J.~D., {Dahal}, S., {Datta}, R., {Denis}, K.~L., {D{\"u}nner}, R., {Espinoza}, F., {Essinger-Hileman}, T., {Flux{\'a} Rojas}, P., {Harrington}, K., {Iuliano}, J., {Karakla}, J., {Marriage}, T.~A., {Miller}, N.~J., {Novack}, S., {N{\'u}{\~n}ez}, C., {Petroff}, M.~A., {Reeves}, R.~A., {Rostem}, K., {Shi}, R., {Valle}, D. A.~N., {Watts}, D.~J., {Weiland}, J.~L., {Wollack}, E.~J., {Xu}, Z., {Zeng}, L., and {Class Collaboration}, ``{CLASS Data Pipeline and Maps for 40 GHz Observations through 2022},'' {\em \apj}~{\bf 956},  77 (Oct. 2023).

\bibitem{shi24}
{Shi}, R., {Appel}, J.~W., {Bennett}, C.~L., {Bustos}, R., {Chuss}, D.~T., {Dahal}, S., {Denes Couto}, J., {Eimer}, J.~R., {Essinger-Hileman}, T., {Harrington}, K., {Iuliano}, J., {Li}, Y., {Marriage}, T.~A., {Petroff}, M.~A., {Rostem}, K., {Song}, Z., {Valle}, D. A.~N., {Watts}, D.~J., {Weiland}, J.~L., {Wollack}, E.~J., and {Xu}, Z., ``{Sensitivity-Improved Polarization Maps at 40 GHz with CLASS and WMAP data},'' {\em arXiv e-prints} ,  arXiv:2404.17567 (Apr. 2024).

\bibitem{eimer2022spie}
{Eimer}, J.~R., {Brewer}, M.~K., {Chuss}, D.~T., {Karakla}, J., {Shi}, R., {Appel}, J.~W., {Bennett}, C.~L., {Cleary}, J., {Dahal}, S., {Datta}, R., {Essinger-Hileman}, T., {Marriage}, T.~A., {N{\'u}{\~n}ez}, C., {Petroff}, M.~A., {Watts}, D.~J., {Wollack}, E.~J., and {Xu}, Z., ``{Construction of a large diameter reflective half-wave plate modulator for millimeter wave applications},'' in [{\em Millimeter, Submillimeter, and Far-Infrared Detectors and Instrumentation for Astronomy XI}{\nolinebreak\hspace{0.1em}]},  {Zmuidzinas}, J. and {Gao}, J.-R., eds., {\em Society of Photo-Optical Instrumentation Engineers (SPIE) Conference Series} {\bf 12190},  121901N (Aug. 2022).

\bibitem{abs14}
{Kusaka}, A., {Essinger-Hileman}, T., {Appel}, J.~W., {Gallardo}, P., {Irwin}, K.~D., {Jarosik}, N., {Nolta}, M.~R., {Page}, L.~A., {Parker}, L.~P., {Raghunathan}, S., {Sievers}, J.~L., {Simon}, S.~M., {Staggs}, S.~T., and {Visnjic}, K., ``{Modulation of cosmic microwave background polarization with a warm rapidly rotating half-wave plate on the Atacama B-Mode Search instrument},'' {\em Review of Scientific Instruments}~{\bf 85},  024501 (Feb. 2014).

\bibitem{simons19whitepaper}
{Simons Observatory Collaboration}, {Ade}, P., {Aguirre}, J., {Ahmed}, Z., {Aiola}, S., {Ali}, A., {Alonso}, D., {Alvarez}, M.~A., {Arnold}, K., {Ashton}, P., {Austermann}, J., {Awan}, H., {Baccigalupi}, C., {Baildon}, T., {Barron}, D., {Battaglia}, N., {Battye}, R., {Baxter}, E., {Bazarko}, A., {Beall}, J.~A., {Bean}, R., {Beck}, D., {Beckman}, S., {Beringue}, B., {Bianchini}, F., {Boada}, S., {Boettger}, D., {Bond}, J.~R., {Borrill}, J., {Brown}, M.~L., {Bruno}, S.~M., {Bryan}, S., {Calabrese}, E., {Calafut}, V., {Calisse}, P., {Carron}, J., {Challinor}, A., {Chesmore}, G., {Chinone}, Y., {Chluba}, J., {Cho}, H.-M.~S., {Choi}, S., {Coppi}, G., {Cothard}, N.~F., {Coughlin}, K., {Crichton}, D., {Crowley}, K.~D., {Crowley}, K.~T., {Cukierman}, A., {D'Ewart}, J.~M., {D{\"u}nner}, R., {de Haan}, T., {Devlin}, M., {Dicker}, S., {Didier}, J., {Dobbs}, M., {Dober}, B., {Duell}, C.~J., {Duff}, S., {Duivenvoorden}, A., {Dunkley}, J., {Dusatko}, J., {Errard}, J., {Fabbian}, G., {Feeney}, S., {Ferraro}, S.,
  {Flux{\`a}}, P., {Freese}, K., {Frisch}, J.~C., {Frolov}, A., {Fuller}, G., {Fuzia}, B., {Galitzki}, N., {Gallardo}, P.~A., {Tomas Galvez Ghersi}, J., {Gao}, J., {Gawiser}, E., {Gerbino}, M., {Gluscevic}, V., {Goeckner-Wald}, N., {Golec}, J., {Gordon}, S., {Gralla}, M., {Green}, D., {Grigorian}, A., {Groh}, J., {Groppi}, C., {Guan}, Y., {Gudmundsson}, J.~E., {Han}, D., {Hargrave}, P., {Hasegawa}, M., {Hasselfield}, M., {Hattori}, M., {Haynes}, V., {Hazumi}, M., {He}, Y., {Healy}, E., {Henderson}, S.~W., {Hervias-Caimapo}, C., {Hill}, C.~A., {Hill}, J.~C., {Hilton}, G., {Hilton}, M., {Hincks}, A.~D., {Hinshaw}, G., {Hlo{\v{z}}ek}, R., {Ho}, S., {Ho}, S.-P.~P., {Howe}, L., {Huang}, Z., {Hubmayr}, J., {Huffenberger}, K., {Hughes}, J.~P., {Ijjas}, A., {Ikape}, M., {Irwin}, K., {Jaffe}, A.~H., {Jain}, B., {Jeong}, O., {Kaneko}, D., {Karpel}, E.~D., {Katayama}, N., {Keating}, B., {Kernasovskiy}, S.~S., {Keskitalo}, R., {Kisner}, T., {Kiuchi}, K., {Klein}, J., {Knowles}, K., {Koopman}, B., {Kosowsky}, A.,
  {Krachmalnicoff}, N., {Kuenstner}, S.~E., {Kuo}, C.-L., {Kusaka}, A., {Lashner}, J., {Lee}, A., {Lee}, E., {Leon}, D., {Leung}, J. S.~Y., {Lewis}, A., {Li}, Y., {Li}, Z., {Limon}, M., {Linder}, E., {Lopez-Caraballo}, C., {Louis}, T., {Lowry}, L., {Lungu}, M., {Madhavacheril}, M., {Mak}, D., {Maldonado}, F., {Mani}, H., {Mates}, B., {Matsuda}, F., {Maurin}, L., {Mauskopf}, P., {May}, A., {McCallum}, N., {McKenney}, C., {McMahon}, J., {Meerburg}, P.~D., {Meyers}, J., {Miller}, A., {Mirmelstein}, M., {Moodley}, K., {Munchmeyer}, M., {Munson}, C., {Naess}, S., {Nati}, F., {Navaroli}, M., {Newburgh}, L., {Nguyen}, H.~N., {Niemack}, M., {Nishino}, H., {Orlowski-Scherer}, J., {Page}, L., {Partridge}, B., {Peloton}, J., {Perrotta}, F., {Piccirillo}, L., {Pisano}, G., {Poletti}, D., {Puddu}, R., {Puglisi}, G., {Raum}, C., {Reichardt}, C.~L., {Remazeilles}, M., {Rephaeli}, Y., {Riechers}, D., {Rojas}, F., {Roy}, A., {Sadeh}, S., {Sakurai}, Y., {Salatino}, M., {Sathyanarayana Rao}, M., {Schaan}, E., {Schmittfull}, M.,
  {Sehgal}, N., {Seibert}, J., {Seljak}, U., {Sherwin}, B., {Shimon}, M., {Sierra}, C., {Sievers}, J., {Sikhosana}, P., {Silva-Feaver}, M., {Simon}, S.~M., {Sinclair}, A., {Siritanasak}, P., {Smith}, K., {Smith}, S.~R., {Spergel}, D., {Staggs}, S.~T., {Stein}, G., {Stevens}, J.~R., {Stompor}, R., {Suzuki}, A., {Tajima}, O., {Takakura}, S., {Teply}, G., {Thomas}, D.~B., {Thorne}, B., {Thornton}, R., {Trac}, H., {Tsai}, C., {Tucker}, C., {Ullom}, J., {Vagnozzi}, S., {van Engelen}, A., {Van Lanen}, J., {Van Winkle}, D.~D., {Vavagiakis}, E.~M., {Verg{\`e}s}, C., {Vissers}, M., {Wagoner}, K., {Walker}, S., {Ward}, J., {Westbrook}, B., {Whitehorn}, N., {Williams}, J., {Williams}, J., {Wollack}, E.~J., {Xu}, Z., {Yu}, B., {Yu}, C., {Zago}, F., {Zhang}, H., and {Zhu}, N., ``{The Simons Observatory: science goals and forecasts},'' {\em \jcap}~{\bf 2019},  056 (Feb. 2019).

\bibitem{litebird22}
{LiteBIRD Collaboration}, {Allys}, E., {Arnold}, K., {Aumont}, J., {Aurlien}, R., {Azzoni}, S., {Baccigalupi}, C., {Banday}, A.~J., {Banerji}, R., {Barreiro}, R.~B., {Bartolo}, N., {Bautista}, L., {Beck}, D., {Beckman}, S., {Bersanelli}, M., {Boulanger}, F., {Brilenkov}, M., {Bucher}, M., {Calabrese}, E., {Campeti}, P., {Carones}, A., {Casas}, F.~J., {Catalano}, A., {Chan}, V., {Cheung}, K., {Chinone}, Y., {Clark}, S.~E., {Columbro}, F., {D'Alessandro}, G., {de Bernardis}, P., {de Haan}, T., {de la Hoz}, E., {De Petris}, M., {Torre}, S.~D., {Diego-Palazuelos}, P., {Dobbs}, M., {Dotani}, T., {Duval}, J.~M., {Elleflot}, T., {Eriksen}, H.~K., {Errard}, J., {Essinger-Hileman}, T., {Finelli}, F., {Flauger}, R., {Franceschet}, C., {Fuskeland}, U., {Galloway}, M., {Ganga}, K., {Gerbino}, M., {Gervasi}, M., {G{\'e}nova-Santos}, R.~T., {Ghigna}, T., {Giardiello}, S., {Gjerl{\o}w}, E., {Grain}, J., {Grupp}, F., {Gruppuso}, A., {Gudmundsson}, J.~E., {Halverson}, N.~W., {Hargrave}, P., {Hasebe}, T., {Hasegawa}, M.,
  {Hazumi}, M., {Henrot-Versill{\'e}}, S., {Hensley}, B., {Hergt}, L.~T., {Herman}, D., {Hivon}, E., {Hlozek}, R.~A., {Hornsby}, A.~L., {Hoshino}, Y., {Hubmayr}, J., {Ichiki}, K., {Iida}, T., {Imada}, H., {Ishino}, H., {Jaehnig}, G., {Katayama}, N., {Kato}, A., {Keskitalo}, R., {Kisner}, T., {Kobayashi}, Y., {Kogut}, A., {Kohri}, K., {Komatsu}, E., {Komatsu}, K., {Konishi}, K., {Krachmalnicoff}, N., {Kuo}, C.~L., {Lamagna}, L., {Lattanzi}, M., {Lee}, A.~T., {Leloup}, C., {Levrier}, F., {Linder}, E., {Luzzi}, G., {Macias-Perez}, J., {Maciaszek}, T., {Maffei}, B., {Maino}, D., {Mandelli}, S., {Mart{\'\i}nez-Gonz{\'a}lez}, E., {Masi}, S., {Massa}, M., {Matarrese}, S., {Matsuda}, F.~T., {Matsumura}, T., {Mele}, L., {Migliaccio}, M., {Minami}, Y., {Moggi}, A., {Montgomery}, J., {Montier}, L., {Morgante}, G., {Mot}, B., {Nagano}, Y., {Nagasaki}, T., {Nagata}, R., {Nakano}, R., {Namikawa}, T., {Nati}, F., {Natoli}, P., {Nerval}, S., {Noviello}, F., {Odagiri}, K., {Oguri}, S., {Ohsaki}, H., {Pagano}, L., {Paiella},
  A., {Paoletti}, D., {Passerini}, A., {Patanchon}, G., {Piacentini}, F., {Piat}, M., {Pisano}, G., {Polenta}, G., {Poletti}, D., {Prouv{\'e}}, T., {Puglisi}, G., {Rambaud}, D., {Raum}, C., {Realini}, S., {Reinecke}, M., {Remazeilles}, M., {Ritacco}, A., {Roudil}, G., {Rubino-Martin}, J.~A., {Russell}, M., {Sakurai}, H., {Sakurai}, Y., {Sasaki}, M., {Scott}, D., {Sekimoto}, Y., {Shinozaki}, K., {Shiraishi}, M., {Shirron}, P., {Signorelli}, G., {Spinella}, F., {Stever}, S., {Stompor}, R., {Sugiyama}, S., {Sullivan}, R.~M., {Suzuki}, A., {Svalheim}, T.~L., {Switzer}, E., {Takaku}, R., {Takakura}, H., {Takase}, Y., {Tartari}, A., {Terao}, Y., {Thermeau}, J., {Thommesen}, H., {Thompson}, K.~L., {Tomasi}, M., {Tominaga}, M., {Tristram}, M., {Tsuji}, M., {Tsujimoto}, M., {Vacher}, L., {Vielva}, P., {Vittorio}, N., {Wang}, W., {Watanuki}, K., {Wehus}, I.~K., {Weller}, J., {Westbrook}, B., {Wilms}, J., {Winter}, B., {Wollack}, E.~J., {Yumoto}, J., {Zannoni}, M., and {Collaboration LiteB I R D}, ``{Probing cosmic
  inflation with the LiteBIRD cosmic microwave background polarization survey},'' {\em Progress of Theoretical and Experimental Physics}~{\bf 2023},  042F01 (Apr. 2023).

\bibitem{pisano2016multi}
{Pisano}, G., {Maffei}, B., {Ade}, P. A.~R., {de Bernardis}, P., {de Maagt}, P., {Ellison}, B., {Henry}, M., {Ng}, M.~W., {Schortt}, B., and {Tucker}, C., ``{Multi-octave metamaterial reflective half-wave plate for millimeter and sub-millimeter wave applications},'' {\em \ao}~{\bf 55},  10255 (Dec. 2016).

\bibitem{harrington18spie}
{Harrington}, K., {Eimer}, J., {Chuss}, D.~T., {Petroff}, M., {Cleary}, J., {DeGeorge}, M., {Grunberg}, T.~W., {Ali}, A., {Appel}, J.~W., {Bennett}, C.~L., {Brewer}, M., {Bustos}, R., {Chan}, M., {Couto}, J., {Dahal}, S., {Denis}, K., {D{\"u}nner}, R., {Essinger-Hileman}, T., {Fluxa}, P., {Halpern}, M., {Hilton}, G., {Hinshaw}, G.~F., {Hubmayr}, J., {Iuliano}, J., {Karakla}, J., {Marriage}, T., {McMahon}, J., {Miller}, N.~J., {Nu{\~n}ez}, C., {Padilla}, I.~L., {Palma}, G., {Parker}, L., {Pradenas Marquez}, B., {Reeves}, R., {Reintsema}, C., {Rostem}, K., {Augusto Nunes Valle}, D., {Van Engelhoven}, T., {Wang}, B., {Wang}, Q., {Watts}, D., {Weiland}, J., {Wollack}, E., {Xu}, Z., {Yan}, Z., and {Zeng}, L., ``{Variable-delay polarization modulators for the CLASS telescopes},'' in [{\em Society of Photo-Optical Instrumentation Engineers (SPIE) Conference Series}{\nolinebreak\hspace{0.1em}]},  {\em Society of Photo-Optical Instrumentation Engineers (SPIE) Conference Series} {\bf 10708},  107082M (July 2018).

\bibitem{dahal22}
{Dahal}, S., {Appel}, J.~W., {Datta}, R., {Brewer}, M.~K., {Ali}, A., {Bennett}, C.~L., {Bustos}, R., {Chan}, M., {Chuss}, D.~T., {Cleary}, J., {Couto}, J.~D., {Denis}, K.~L., {D{\"u}nner}, R., {Eimer}, J., {Espinoza}, F., {Essinger-Hileman}, T., {Golec}, J.~E., {Harrington}, K., {Helson}, K., {Iuliano}, J., {Karakla}, J., {Li}, Y., {Marriage}, T.~A., {McMahon}, J.~J., {Miller}, N.~J., {Novack}, S., {N{\'u}{\~n}ez}, C., {Osumi}, K., {Padilla}, I.~L., {Palma}, G.~A., {Parker}, L., {Petroff}, M.~A., {Reeves}, R., {Rhoades}, G., {Rostem}, K., {Valle}, D. A.~N., {Watts}, D.~J., {Weiland}, J.~L., {Wollack}, E.~J., and {Xu}, Z., ``{Four-year Cosmology Large Angular Scale Surveyor (CLASS) Observations: On-sky Receiver Performance at 40, 90, 150, and 220 GHz Frequency Bands},'' {\em \apj}~{\bf 926},  33 (Feb. 2022).

\bibitem{novak1989a}
{Novak}, G., {Sundwall}, J.~L., and {Pernic}, R.~J., ``{Far infrared polarizing grids for use at cryogenic temperatures},'' {\em \ao}~{\bf 28},  3425--3427 (Aug. 1989).

\bibitem{Voellmer2008}
{Voellmer}, G.~M., {Bennett}, C., {Chuss}, D.~T., {Eimer}, J., {Hui}, H., {Moseley}, S.~H., {Novak}, G., {Wollack}, E.~J., and {Zeng}, L., ``{A large free-standing wire grid for microwave variable-delay polarization modulation},'' in [{\em Ground-based and Airborne Instrumentation for Astronomy II}{\nolinebreak\hspace{0.1em}]},  {McLean}, I.~S. and {Casali}, M.~M., eds., {\em \procspie}~{\bf 7014},  70142A (July 2008).

\bibitem{Chuss2014}
{Chuss}, D.~T., {Eimer}, J.~R., {Fixsen}, D.~J., {Hinderks}, J., {Kogut}, A.~J., {Lazear}, J., {Mirel}, P., {Switzer}, E., {Voellmer}, G.~M., and {Wollack}, E.~J., ``{Variable-delay polarization modulators for cryogenic millimeter-wave applications},'' {\em Review of Scientific Instruments}~{\bf 85},  064501 (June 2014).

\bibitem{datta2023class}
{Datta}, R., {Brewer}, M.~K., {Denes Couto}, J., {Eimer}, J., {Li}, Y., {Xu}, Z., {Ali}, A., {Appel}, J.~W., {Bennett}, C.~L., {Bustos}, R., {Chuss}, D.~T., {Cleary}, J., {Dahal}, S., {Espinoza}, F., {Essinger-Hileman}, T., {Flux{\'a}}, P., {Harrington}, K., {Helson}, K., {Iuliano}, J., {Karakla}, J., {Marriage}, T.~A., {Novack}, S., {N{\'u}{\~n}ez}, C., {Padilla}, I.~L., {Parker}, L., {Petroff}, M.~A., {Reeves}, R., {Rostem}, K., {Shi}, R., {Valle}, D. A.~N., {Watts}, D.~J., {Weiland}, J.~L., {Wollack}, E.~J., and {Zeng}, L., ``{Cosmology Large Angular Scale Surveyor (CLASS): 90 GHz Telescope Pointing, Beam Profile, Window Function, and Polarization Performance},'' {\em arXiv e-prints} ,  arXiv:2308.13309 (Aug. 2023).

\bibitem{komatsu2020design}
{Komatsu}, K., {Ishino}, H., {Katayama}, N., {Matsumura}, T., {Sakurai}, Y., and {Takaku}, R., ``{Design of the frequency independent optic axis of the Pancharatnam base achromatic half-wave plate for CMB polarization experiment},'' in [{\em Millimeter, Submillimeter, and Far-Infrared Detectors and Instrumentation for Astronomy X}{\nolinebreak\hspace{0.1em}]},  {Zmuidzinas}, J. and {Gao}, J.-R., eds., {\em \procspie}~{\bf 11453},  114534B (Dec. 2020).

\bibitem{adler2024modeling}
{Adler}, A.~E., {Austermann}, J.~E., {Benton}, S.~J., {Duff}, S.~M., {Filippini}, J.~P., {Fraisse}, A.~A., {Gascard}, T., {Gibbs}, S.~M., {Gourapura}, S., {Hubmayr}, J., {Gudmundsson}, J.~E., {Jones}, W.~C., {May}, J.~L., {Nagy}, J.~M., {Okun}, K., {Padilla}, I., {Rooney}, C., {Tartakovsky}, S., and {Vissers}, M.~R., ``{Modeling optical systematics for the Taurus CMB experiment},'' {\em arXiv e-prints} ,  arXiv:2406.11992 (June 2024).

\end{thebibliography}
\bibliographystyle{spiebib} 

\end{document}